\documentclass{jfm}
\usepackage{amsmath}
\usepackage{amssymb}
\usepackage{float}
\usepackage{dcolumn}
\usepackage{bm}
\usepackage{newtxmath}
\usepackage{natbib}
\usepackage{hyperref}

\hypersetup{
   colorlinks = true,
   urlcolor   = blue,
   citecolor  = blue,
   linkcolor  = blue,
}

\title{Cavitation bubble dynamics inside a droplet suspended in a different host fluid}

\author{Shuai Li\aff{1}, 
	Zhesheng Zhao\aff{1}, 
 A-Man Zhang\aff{1}
 \corresp{\email{zhangaman@hrbeu.edu.cn}},
\and Rui Han\aff{2}}

\affiliation{\aff{1}College of Shipbuilding Engineering, Harbin Engineering University, Harbin, China
\aff{2}Heilongjiang Provincial Key Laboratory of Nuclear Power System \& Equipment, Harbin Engineering University, Harbin, China}

\begin{document}
\maketitle

\begin{abstract}

In this paper, we present a theoretical, experimental, and numerical study of the dynamics of cavitation bubbles inside a droplet suspended in another host fluid. On the theoretical side, we provided a modified Rayleigh collapse time and natural frequency for spherical bubbles in our particular context, characterized by the density ratio between the two liquids and the bubble-to-droplet size ratio. Regarding the experimental aspect, experiments were carried out for laser-induced cavitation bubbles inside oil-in-water (O/W) or water-in-oil (W/O) droplets. Two distinct fluid-mixing mechanisms were unveiled in the two systems, respectively. In the case of O/W droplets, a liquid jet emerges around the end of the bubble collapse phase, effectively penetrating the droplet interface. We offer a detailed analysis of the criteria governing jet penetration, involving the standoff parameter and impact velocity of the bubble jet on the droplet surface. Conversely, in the scenario involving W/O droplets, the bubble traverses the droplet interior, inducing global motion and eventually leading to droplet pinch-off when the local Weber number exceeds a critical value. This phenomenon is elucidated through the equilibrium between interfacial and kinetic energies. Lastly, our boundary integral model faithfully reproduces the essential physics of nonspherical bubble dynamics observed in the experiments. We conduct a parametric study spanning a wide parameter space to investigate bubble-droplet interactions. The insights from this study could serve as a valuable reference for practical applications in the field of ultrasonic emulsification, pharmacy, etc.

\end{abstract}



\section{Introduction}
\label{sec:intro}

The dynamics of cavitation bubbles inside liquid droplets have been attracting increasing interest in the scientific community. This is mainly because confined boundaries can lead to surprisingly rich dynamics, and the phenomena have many important applications, such as drop atomization/fragmentation \citep{Gonzalez}, nanolithography \citep{Banine}, ultrasonic emulsification \citep{Stepisnik, Orthaber}, and liquid chromatography \citep{Janzen}, to name a few. A better fundamental understanding of the interaction between cavitation bubbles and droplets is of key importance for the development of the aforementioned applications.

Bubble dynamics in an infinite/semi-infinite liquid domain have been widely and extensively studied in the literature, including interactions with rigid walls \citep{Hsiao2014,Brujan2018,ZengQ2022,Saini2022,Zhang2023a}, a water-air free surface \citep{Kang2019,saade2021,Bempedelis2021,Cerbus2022}, elastic boundaries \citep{Brujan01,Klaseboer04}, fluid-fluid interfaces \citep{Klaseboer-CM2004,Orthaber}, adjacent bubbles \citep{Tomita17,Luo2021}, suspended particles \citep{Poulain15,Ren2022}, and more. Of particular interest are the bubble collapse patterns and jetting behaviors in different situations. However, bubble dynamics inside a liquid droplet exhibit striking differences. For instance, \citet{Obreschkow} conducted experimental and theoretical investigation of cavitation bubble dynamics within a centimeter-sized quasi-spherical water droplet. They found that the typical crown structure surrounding the spike on flat surfaces \citep{saade2021} is absent on the spherical droplet surface. \citet{Gonzalez} experimentally unveiled three distinct fragmentation regimes (atomization, sheet formation, and coarse fragmentation) in levitated droplets exposed to a laser-induced cavitation bubble. The same group \citep{zeng-drop} performed numerical studies on the jetting phenomena from the droplet surface (the coarse fragmentation regime) and proved that a theoretical model grounded in the spherical Rayleigh-Taylor instability can  predict the onset of jetting on the droplet surface. \citet{WangJZ2021} experimentally and theoretically studied the interfacial instability of cylindrical (two-dimensional) water droplets driven by a laser-induced cavitation bubble positioned at the center. They found that the slight perturbation on the droplet surface easily grows, and the bubble interior may connect to ambient air under certain conditions. Due to the large density ratio between the droplet and surrounding air, the air flow basically plays a minor role but provides a constant pressure boundary.

\begin{figure}
	\centering\includegraphics[width=12cm]{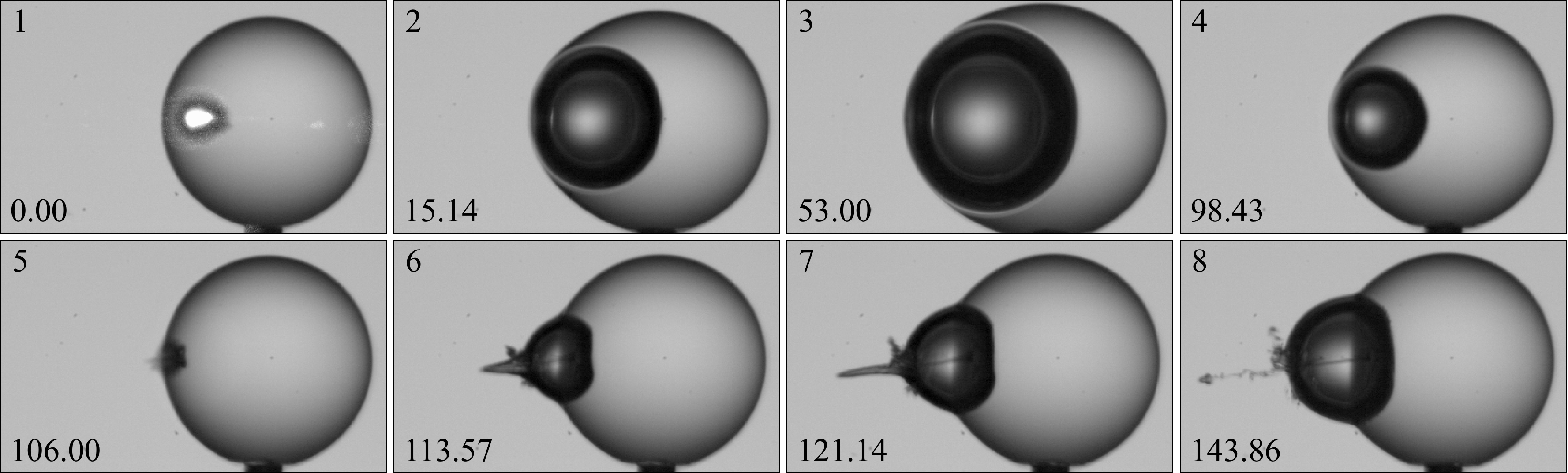}
	\caption{Experimental observation of a laser-induced cavitation bubble inside a sunflower oil droplet (density: 914 $\mathrm{kg} / \mathrm{m}^{3}$) suspended in water. The oil droplet’s initial radius is 0.69 mm, and the bubble is generated using the laser focusing method, reaching a maximum radius of approximately 0.58 mm. Dimensional times are indicated in each frame’s lower left corners (Unit: \textmu s). Each frame’s horizontal width measures 2.5 mm. The distance from the initial bubble center to the nearest droplet surface is about 0.27 mm. Detailed information about the experimental set-up is provided in later sections.}\label{Fig:1}
\end{figure}

We can learn from the forementioned studies that the spherical air-water interface driven by an oscillating cavitation bubble is generally associated with Rayleigh–Taylor instabilities \citep{zeng-drop,Rossello}. Specifically, fragmentation/atomization of a droplet can be easily achieved by using a cavitation bubble \citep{Gonzalez,Liang2020,Klein2020}. Intuitively, we would expect a similar mechanism in the physical process of ultrasonic emulsification, which is widely used in industries where water and oil are simultaneously used to produce nano-emulsions where bubble-droplet interaction plays a key role \citep{Stepisnik, Raman2022JFM,HanH2023}. However, this problem differs from droplets levitated in air as the droplet is suspended in another host fluid with comparable density. As a result, the Rayleigh-Taylor instability of the interface and subsequent droplet fragmentation are rarely observed. This is exemplified by experimental observations involving a relatively large cavitation bubble within a sunflower oil droplet (density: 914 kg/m³) suspended in water, as illustrated in Figure \ref{Fig:1}. One can also examine this using the spherical Rayleigh-Taylor instability theory \citep{Plesset1954}. Based on the observations from frames 5-8 of Figure \ref{Fig:1}, a dominant mechanism for fluid mixing is identified in this O/W system. It involves the high-speed jet generated by the bubble impacting the droplet surface, which effectively transports oil droplets into the surrounding water.

Since it is difficult to observe the microscopic phenomena during the transient interaction of micro-droplets with acoustic bubbles, some preliminary studies on the interaction between cavitation bubbles and a flat water-oil interface can shed light on the physics of bubbles in binary immiscible fluids. The direction of the bubble jet is generally from the lighter liquid to the denser liquid if the gravity effect is weak \citep{Orthaber}, which is a crucial mechanism of fluid mixing in ultrasonic emulsification. \citet{Han2022} discovered that the pinch-off of an interface jet occurs long after the bubble dynamics stage, representing another mechanism for fluid mixing. These studies can be extended to scenarios involving a large droplet and relatively small bubbles. It would be intriguing to investigate the impact of a highly curved droplet interface on the dynamics of adjacent bubbles. This is a crucial process in the droplet breakup during ultrasonic emulsification \citep{Orthaber,Yamamoto2021,Udepurkar,HanH2023}.

In practical ultrasonic emulsification, ultrasonic waves propagate throughout the liquid medium, resulting in the formation of cavitation bubbles in both the water and oil phases \citep{Wu2023}. As a result, bubble nucleation can take place both externally and within droplets. A recent research by \citet{Raman2022JFM} explored the interaction between a submillimeter-sized water droplet and a nearby laser-induced cavitation bubble submerged in silicone oil. In our study, we investigate a distinct scenario where cavitation bubbles are initially formed within droplets. Notably, \citet{Stepisnik} highlighted a relevant aspect during the emulsification of oil-water systems. As oil droplets enter the water phase, they carry numerous cavitation nuclei within them. This phenomenon substantially increases the probability of bubble expansion and subsequent collapse within these oil droplets. Such insights emphasize the practical significance of the issues we are investigating. The experimental and numerical results we present reveal remarkable differences from previous findings \citep{Raman2022JFM, Raman2022}. For instance, we identify a novel mechanism for fluid mixing within the W/O system. Specifically, the bubble is repelled by the adjacent droplet surface, then travels a considerable distance within the droplet, ultimately pushing against the opposite side of the droplet surface. This dynamic process can result in the pinch-off of the droplet when the local Weber number surpasses a critical threshold. Our research encompasses investigations in both W/O and O/W systems, each of which holds substantial implications for understanding the mechanisms underlying droplet breakup.

In this study, we first present a modified Rayleigh collapse time and the natural frequency of spherical bubbles in our particular context. Subsequently, we conducted an extensive series of experiments involving laser-induced cavitation bubbles and boundary integral simulations. Our objective is to reveal the dependence of the bubble jetting behaviors and the associated droplet evolution on the governing parameters, including the nondimensional standoff parameter, the density ratio between the two liquids, and the bubble-to-droplet size ratio.  Particularly, we offer an in-depth analysis of the criteria governing jet penetration and droplet pinch-off.


The following is the structure of this work. First, \S\  \ref{sec:method} introduces the methodology of this study. In \S\  \ref{sect: Exp.}, the general physical phenomena related to bubble initiation in both O/W  and W/O droplets are discussed. In \S\  \ref{sect: spherical}, we examine the dynamics of a spherical bubble inside the droplet. In \S\  \ref{sect: nonspherical}, we provide a quantitative discussion of the dependence of nonspherical bubble dynamics on governing parameters. In \S\  \ref{sect:discussion}, we discuss two mechanisms responsible for fluid mixing in O/W and W/O systems. Finally, we present our conclusions in \S\  \ref{sect:conclusion}.




\section{Methodology}
\label{sec:method}
\subsection {Theoretical model}
\label{sec:method1}
On the theoretical side, we are considering a spherical, inertial, oscillating bubble located at the center of a spherical droplet (referred to as fluid 1) that is surrounded by the host liquid (referred to as fluid 2). This is illustrated in Figure \ref{Fig:sketch} (a). The bubble dynamic equation in our particular context was originally formulated by \cite{Raman2022JFM}. They reached this equation through a modification of a model for bubble dynamics with a linear elastic shell surrounded by a viscous fluid, as initially described in \cite{Church}. In order to improve clarity and facilitate a more intuitive comprehension of the equation, particularly for researchers who may not have extensive expertise in this field, we offer an alternative derivation based on the Laplace equation and the Bernoulli equation as follows.

The bubble and droplet radii are denoted by $R_b$ and $R_d$, respectively. The densities of fluid 1 and fluid 2 are denoted by $\rho_1$ and $\rho_2$, respectively. The flow dynamics in both fluids are governed by the Laplace equation
\begin{equation}
\nabla^2{\varphi_i}=0\ \ (i = 1, 2), \ \   \label{Equation:Laplace}
\end{equation}
where  $\varphi$ represents the velocity potential, and the subscript `$i$' represents the fluid type. The velocity and velocity potential in the entire flow field can be expressed in unified forms, irrespective of the fluid type, as follows:

\begin{figure}
	\centering\includegraphics[width=12cm]{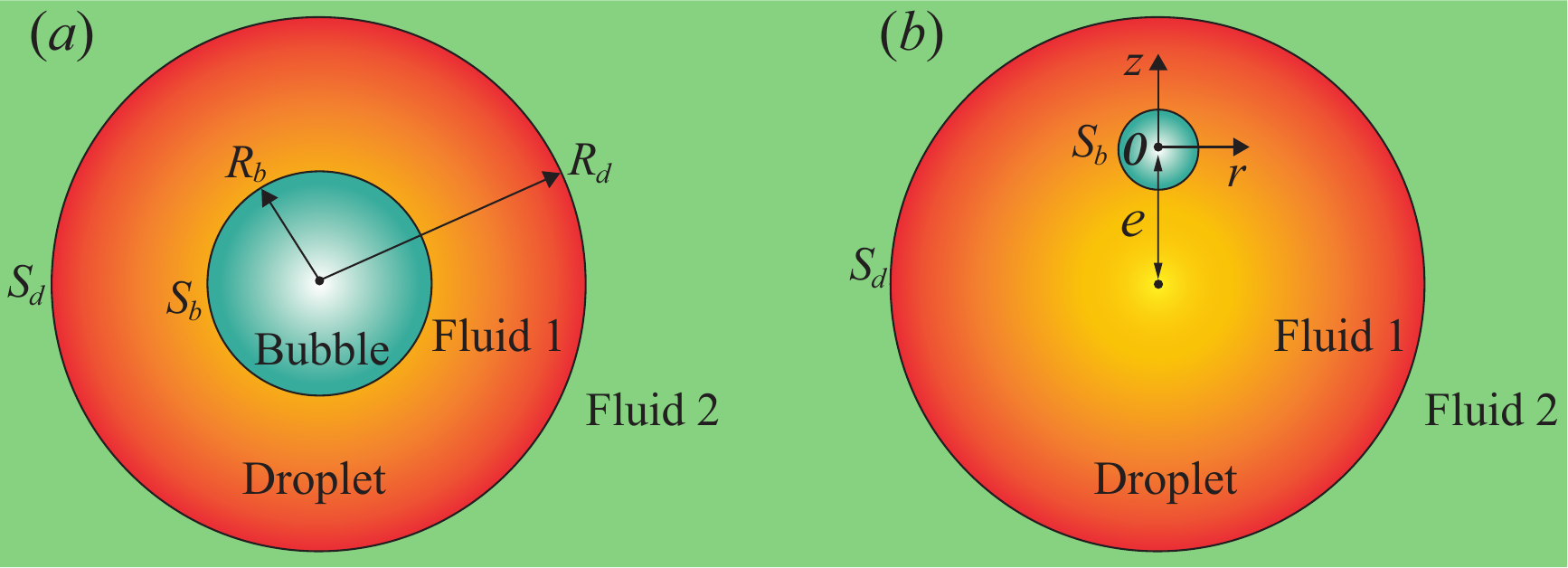}
	\caption{Sketch illustrating the physical problem of bubble dynamics within an initially spherical droplet (referred to as fluid 1) surrounded by the host liquid (referred to as fluid 2).	(a) The bubble is situated at the center of the droplet, with $R_b$ and $R_d$ denoting the radii of the bubble and droplet, respectively. The surfaces of the bubble and droplet are represented by $S_b$ and $S_d$, respectively. (b) The bubble is not positioned centrally within the droplet, and the eccentricity between the bubble and the droplet centroids is denoted by $e$.}\label{Fig:sketch}
\end{figure}

\begin{equation}
\nabla{\varphi_i}=\frac{\dot{R_b} R_b^2}{r^2}, \ \ \varphi_i=-\frac{\dot{R_b}R_b^2}{r} \ \ (i = 1, 2),  \label{Equation:u-phi}
\end{equation}
where the dot presents the time derivative, and $r$ denotes the radial position in the flow field from the center of the bubble.

Considering the effect of surface tension, we can express the relationship between the gas pressure inside the bubble ($P_g$) and the liquid pressure on the bubble surface ($P_{b1}$) as follows:

\begin{equation} 
P_g=P_{b1}+\frac{2\sigma_1}{R_b},\label{Equation:Pg}
\end{equation}
where $\sigma_1$ is the surface tension coefficient for the bubble surface. Similarly, the relation between the pressures just inside and outside the droplet surface, denoted by $P_{d1}$ and $P_{d2}$ respectively, can be expressed as

\begin{equation} 
P_{d1}=P_{d2}+\frac{2\sigma_2}{R_d},\label{Equation:Pd}
\end{equation}
where $\sigma_2$ presents the surface tension coefficient of the fluid-fluid interface.

The unsteady Bernoulli equation is applicable to both fluids. As a result, we can express the dynamic boundary conditions on the bubble surface and the fluid-fluid interface as follows:

\begin{equation} 
{\left.\frac{{\partial\varphi_1}}{{\partial}t}\right|_{r=R_b}} +\frac{1}{2}{\dot{R_b}^2}+\frac{P_{b1}}{\rho_1}={\left.\frac{{\partial\varphi_1}}{{\partial}t}\right|_{r=R_d}}+\frac{1}{2}({\frac{\dot{R_b}R_b^2}{R_d^2}})^2+\frac{P_{d1}}{\rho_1} \ \ \ \rm on\ \it S_{b},\label{Equation:Sb}
\end{equation}

\begin{equation} 
{\left.\frac{{\partial\varphi_2}}{{\partial}t}\right|_{r=R_d}} +\frac{1}{2}({\frac{\dot{R_b}R_b^2}{R_d^2}})^2+\frac{P_{d2}}{\rho_2}=\frac{P_\infty}{\rho_2} \ \ \ \rm on\ \it S_{d},\label{Equation:Sd}
\end{equation}
where $P_\infty$ is the hydrostatic pressure at infinity.

The bubble dynamic equation can be obtained from Equations (\ref{Equation:u-phi}-\ref{Equation:Sd}), written as
\begin{equation}
R_b\ddot{R_b} + \frac{3}{2} \dot{R_b}^2+(1-\alpha)\left[ -2\theta \dot{R_b}^2-\theta R_b\ddot{R_b}+\frac{1}{2}\theta^4 \dot{R_b}^2\right]  = \frac{P_g-P_\infty}{\rho_1}-\frac{2\sigma_1}{\rho_1R_b}-\frac{2\theta\sigma_2}{\rho_1R_b},  \label{Equation:ERP}
\end{equation}
where $\theta=R_b/R_d$ is the bubble-to-droplet size ratio that varies with time and $\alpha=\rho_2/\rho_1$ is the density ratio between the two fluids. This equation is an extension form of Rayleigh-Plesset (RP) equation and one of its derivatives. When $\alpha=0$, Equation (\ref{Equation:ERP}) reduces to the form obtained by \citet{Obreschkow}, which describes the bubble dynamics inside a droplet that is levitated in air (without inertia effects). As $\theta$ approaches 0, the RP equation is recovered. Our approach can be extended to derive equations governing the dynamics of bubbles within more fluid layers under spherically symmetric conditions.

To model and match the experimental data, we employ the adiabatic approximation to compute the gas pressure \citep{Klaseboer04,Zeng2020,Han2022,Zhang2023},  given by 
\begin{equation} 
P_g\it=P_{\rm 0}\it\left(\frac{V_{\rm 0}}{V}\rm \right) ^{\it\lambda},\label{Equation:pg}
\end{equation}
where $V$ denotes the bubble volume, $P_0$ the initial bubble pressure, $\lambda= 1.4$ the ratio of the specific heats, and the subscript `0' represents initial quantities. This simplicity can be justified by the large associated Péclet number $\sim O(10^3)$ of the bubbles in our experiments. 

The viscous is neglected here but can be incorporated through the conditions on the normal stresses at the bubble wall and the droplet surface. To estimate the viscous effect, we define the Reynolds number for bubble dynamics as $Re=\rho_1 R_{b, max} \it U/{\mu\rm_1}$, where the characteristic velocity is taken as $U=\sqrt{P_\infty/\rho_1}$, $\mu_1$ is the viscosity of fluid 1, and $R_{b, max}$ is the maximum radius of the bubble. For submillimeter-sized laser-induced bubbles generated in a viscous oil droplet, the Reynolds number can be estimated as $Re\sim10^2$. The deviation of the bubble oscillation period between the results obtained from Equation (\ref{Equation:ERP}) and the equation with viscous terms \citep{Raman2022JFM} is within 2\%. For bubbles generated in a W/O droplet, the variation of the bubble oscillation period due to viscous effects is within 0.2\%. In applications such as ultrasonic emulsification, where the maximum acoustic bubble radius typically falls within the range of 30 to 150 \textmu m \citep{Yamamoto2021,Wu2023,Udepurkar}, it is expected that viscosity would have a more pronounced influence. However, the experimental data with adequate spatio-temporal resolution for such small bubbles is currently lacking. As a result, our investigation is confined to millimeter-sized bubbles, and the examination of viscosity effects on micron-sized bubbles is beyond the scope of this study. Interested readers can refer to the relevant literature \citep{Popinet,Minsier,zeng-drop,Kannan2020,WangQ2022} for more discussion.

Next, we derive a modified Rayleigh collapse time that describes the duration from the bubble's maximum radius $R_{b, \rm max}$ to the point where the bubble is completely filled up under a constant pressure difference $\Delta P = P_\infty - P_g$ \citep{Rayleigh}. Surface tension is neglected here for the sake of clarity. The energy conservation equation can be solved for $\dot{R_b}$ and is expressed as:

\begin{equation} 
\dot{R_b}=-\sqrt{\dfrac{2\Delta P(R^{\prime-3}-1)}{3\rho_1\left[1+\dfrac{(\alpha-1)}{\sqrt[3]{\xi^{-3}R^{\prime-3}+1}}  \right] }},\label{Equation:dotR}
\end{equation}
where $R^{\prime}=R_b/R_{b, max}$ and $\xi=R_{b, max}/R_{d, min}$. In this context, we find it convenient to use a fixed value of the bubble-to-droplet size ratio $\xi$ for a given case. 

Integrating this equation from $R_b=R_{b,  max}$ to $R_b=0$ yields an analytical expression for the bubble collapse time,

\begin{equation} 
T_c=\eta(\alpha,\ \xi) \cdot R_{b, max}\sqrt{\frac{\rho_1}{\Delta P}}
\label{Equation:collapse-time}
\end{equation}
with
\begin{equation}  
\eta(\alpha,\ \xi)=\sqrt{\frac{1}{6}}\int_{0}^{1}\sqrt{1+\frac{\alpha-1}{(\xi^{-3}s^{-1}+1)^{1/3}}}\cdot \frac{{\rm d}s}{s^{1/6}(1-s)^{1/2}},
\label{Equation:factor}
\end{equation}
where $s$ substitutes $R^{\prime 3}$. Note that the modified Rayleigh factor $\eta$ is influenced by both density ratio $\alpha$ and size ratio $\xi$. If $\alpha=1$ or $\xi \to 0$, Equation (\ref{Equation:factor}) reduces to the classic Rayleigh factor $\eta \approx 0.9147$. More quantitative discussion about the influence of $\alpha$ and $\xi$ on bubble collapse time will be given in \S\  \ref{sect: Rayleigh collapse time}.

Finally, if the bubble oscillates with a low-amplitude, the bubble's natural frequency can be obtained in a similar manner as \citet{Minnaert}, given by:
\begin{equation}  
f(\alpha,\ \theta_e)=\frac{1}{2\pi R_e}\sqrt{\cfrac{3\lambda\left(P_{\infty}+\cfrac{2\sigma_1}{R_e}+\cfrac{2\sigma_2\theta_e}{R_e} \right)-\cfrac{2\sigma_1}{R_e}-\cfrac{2\sigma_2\theta_e^4}{R_e} }{\rho_1 \left[1+(\alpha-1)\theta_e \right] }},
\label{Equation:natural frequency}
\end{equation}
where $R_e$ represents the equilibrium radius of the bubble and $\theta_e=R_e/R_d$. This formula would be useful for the community of acoustic bubbles. 

\subsection {Boundary integral method}
\label{sec:method2}

When the bubble is not centered in the droplet, as shown in Figure \ref{Fig:sketch} (b), nonspherical oscillations and jetting behaviors of the bubble can be expected. Therefore, we utilize a well-verified boundary integral (BI) method \citep{Li2020,Yi2021,Han2022,Li_jcp} to investigate the dependence of nonspherical bubble dynamics inside a droplet on the governing parameters. Here, we provide a brief overview of the BI method. We use a cylindrical coordinate system $(r,\theta, z)$, with the origin $O$ located at the center of the bubble and the positive $z$-axis pointing from the initial droplet center towards the bubble center. Through the application of Green's second theorem, the Laplace equation (\ref{Equation:Laplace}) can be transformed into a boundary integral equation, expressed as follows:
\begin{equation}
c(\textbf{\textit{r}})\varphi_i(\textbf{\textit{r}})=\iint_S\left(\frac{\partial \varphi_i(\textbf{\textit{q}})}{\partial\it n}\frac{1}{\left| \textbf{\textit{r}}-\textbf{\textit{q}}\right| }-\varphi_i\it(\textbf{\textit{q}})\frac{\partial}{\partial n}(\frac{\rm 1}{\left| \textbf{\textit{r}}-\textbf{\textit{q}}\right| }) \right) \rm d\it S_i(\textbf{\textit{q}}),\ (\it i\rm=1,2),  \label{Equation:BIE}
\end{equation}
where $c$ is the solid angle, $\textbf{\textit{r}}$ and $\textbf{\textit{q}}$ stand for the control and source points, respectively, $\partial/\partial n$ denotes the normal derivative, and $S$ refers to the droplet surface and the bubble surface when $i = 1$ (flow domain 1), while refers to the droplet interface only when $i = 2$ (flow domain 2).

The dynamic boundary conditions on the bubble surface and the droplet interface are given by

\begin{equation} 
\frac{{\rm D}{\varphi_1}}{{\rm D}t}=\frac{P_{\infty}}{\rho_1}-\frac{P_{g}}{\rho_1}+\frac{1}{2}{|{\nabla \varphi_1}|^2+\frac{\sigma_1 \kappa}{\rho_1}}\ \ \ \rm on\ \it S_{b},\label{Equation:DBC_Sb}
\end{equation}

\begin{equation} 
\frac{{\rm D}{(\varphi_1-\alpha\varphi_2)}}{{\rm D}t}=\frac{1}{2}{|{\nabla \varphi_1}|^2}+\frac{\alpha}{2}{|{\nabla \varphi_2}|^2}-\alpha\nabla\varphi_1\cdot\nabla\varphi_2-\frac{\sigma_2 \kappa}{\it\rho\rm_1} \ \ \ \rm on\ \it S_{d},\label{Equation:dynamic-drop}
\end{equation}
where $\kappa$ denotes the curvature. 

The kinematic boundary condition on both $S_b$ and $S_d$ is expressed as

\begin{equation} 
\frac{{\rm D}{\textbf{\textit{r}}}}{{\rm D}t}=\nabla\varphi_1.\label{Equation:kinematic}
\end{equation}

The boundary integral equations (\ref{Equation:BIE}) are solved for the velocities on bubble and droplet surfaces. Then, we update the velocity potential using Equations (\ref{Equation:DBC_Sb}-\ref{Equation:dynamic-drop}) and the position of the surfaces using Equation (\ref{Equation:kinematic}).  
All simulations are conducted in a nondimensional form, where the maximum radius of the bubble $R_{b, max}$, hydrostatic pressure $P_\infty$, and density of the droplet $\rho_1$ serve as the three fundamental quantities. Four nondimensional variables of the system are given below:

\begin{equation} 
\alpha=\frac{\rho_2}{\rho_1},\  \frac{1}{\xi}=\frac{R_{d, min}}{R_{b, max}}, \   \gamma=\frac{R_{d, min}-e}{R_{b, max}},\   \varepsilon=\frac{P_0}{P_\infty}, \label{Equation:kin}
\end{equation}
where the density ratio $\alpha$  and size ratio $\xi$ have the same definition as in \S\  \ref{sec:method1}, $\gamma$ is the standoff parameter that measures the nondimensional distance from the bubble center to the nearest droplet surface ($e$ is the eccentricity),  $\varepsilon$ is the strength parameter that describes the initial bubble pressure.  As for the toroidal bubble dynamics, we use a vortex ring model \citep{WangQX1996,Curtiss2013,Zhang2015pof,HanL} coupled with the BI method to continue the simulation after jet impact.  

\subsection {Experimental set-up}
\label{sec:method3}

First, experiments were conducted to investigate the dynamics of cavitation bubbles within a water droplet (W/O system) in a tank measuring $100\times100\times100 \rm\ mm^3$ at room temperature ($\sim25 ^{\circ}$C) and atmospheric pressure ($\sim$ 97.1 kPa). The tank was initially filled with sunflower oil (50.6 centistokes) to a depth of 90 mm, with a density of 914 kg/$\rm m^3$. To achieve precise control over the volume and position of millimeter-sized liquid droplets in the oil bulk, we installed an injection syringe fitted with an ultrafine flat-tipped needle (inner diameter 0.08 mm, outer diameter 0.2 mm) on a three-axis mobile platform (precision 0.02mm). This setup allowed us to generate a millimeter-scale droplet hanging at the needle opening. The associated Bond number, defined as $Bo=(\rho_1-\rho_2)gR_{d, min}^2/\sigma_2$, can be estimated to be $O(10^{-2})$. The surface tension coefficient of the water-oil interface $\sigma_2$ is about 0.029 N/m. The difference between the horizontal radius $R_r$ and the vertical radius $R_z$ of the droplet is within 4\%. Therefore, we can assume the droplet is initially spherical. In experiments involving deionized water as the host fluid and sunflower oil as the droplet (O/W system), the needle opening was positioned upward, and the droplet was supported by the needle. While one may concern about the needle's influence on the transient bubble-droplet interaction, the experimental results demonstrated that the impact of the needle on bubble oscillation and jetting behaviors was minimal due to large  difference in size.

\begin{figure}
	\centering
	\includegraphics[width=10cm]{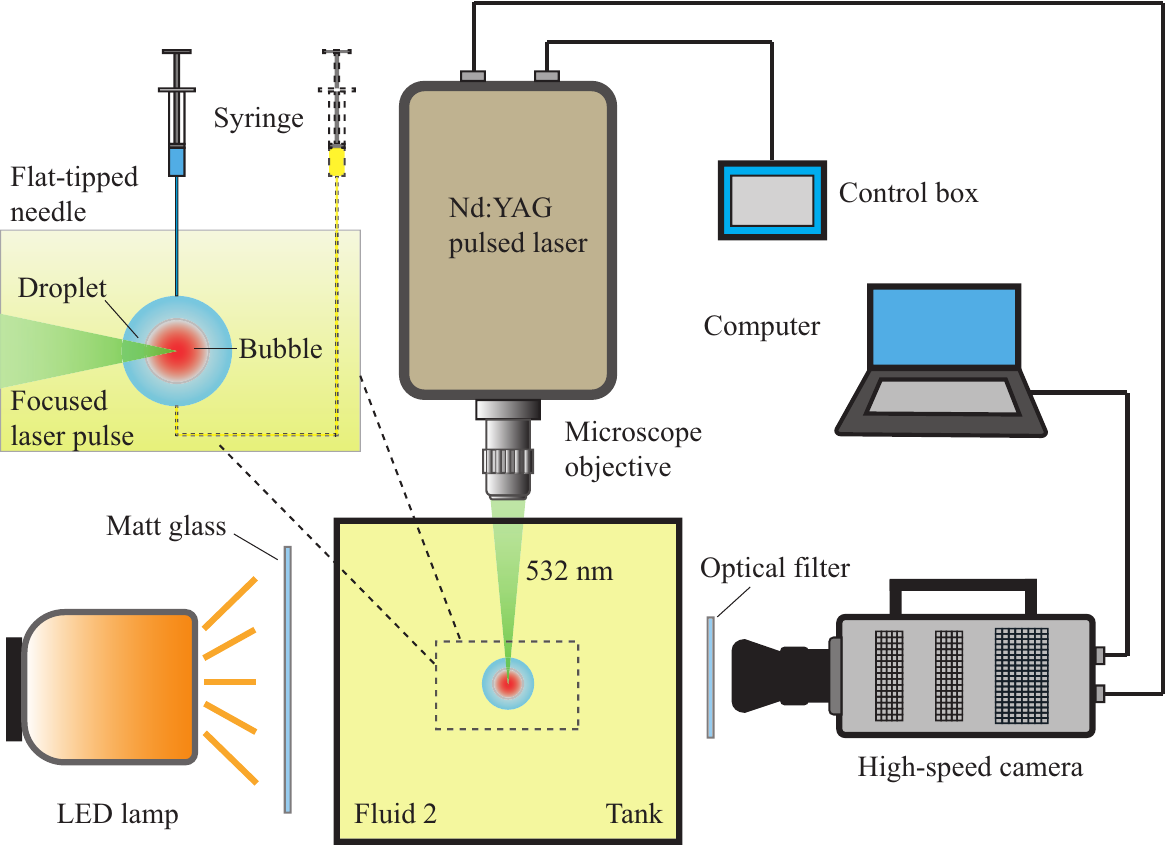}
	\caption{Experimental set-up for cavitation bubble dynamics inside a droplet that surrounded by a different host fluid. A submillimeter-scale cavitation bubble is generated by focusing a pulsed laser inside a droplet that is hanged or hold by a thin flat-tipped needle. }
	\label{schematic diagram}
\end{figure}

The generation of submillimeter-scale cavitation bubbles was achieved using a frequency-doubled Nd:YAG laser (Nimma-900, pulse duration 8 ns, wavelength 532 nm, pulse energy 16$\sim$32 mJ). A microscope objective lens (M Plan Apo L 10$\times$, numerical aperture NA = 0.28) focused the parallel pulsed laser beam inside the droplet. A 30 mm diameter hole was created in the sidewall of the tank, with a 0.25 mm thick sapphire glass embedded to minimize the effects of refraction, thereby enhancing laser focusing. The energy at the focal point exceeded the breakdown threshold of the liquid medium, resulting in transient `avalanche' ionization, forming a high-temperature and high-pressure plasma cavity. Subsequently, it rapidly expanded into a cavitation bubble with a maximum diameter of about 1 mm. 
		
For uniform background illumination, a continuous LED light source (300 W) filtered through matt glass was utilized. To capture both the transient bubble behaviors and the droplet evolutions, a high-speed camera (Phantom V2012) equipped with a macro lens (LAOWA, 100 mm, F2.8) was triggered simultaneously with the laser. The camera recorded the phenomena at a resolution of 256 $\times$ 128 pixels, with 340,000 frames per second and an exposure time of 1 \textmu s. To measure the bubble pulsation period and micro-jet velocity more precisely, we only look at the area near the equator of the droplet-bubble system, enabling us to record at a frame rate of up to 656, 000. Due to the droplet's optical-lens-like effect \citep{Gonzalez}, a direct measurement of the bubble size is questionable. Instead, we calculate the bubble radius based on the changes in droplet volume. We notice that for the cases where bubbles are located at the center of O/W and W/O droplets, the actual sizes of the bubble are 0.94-0.97 and 1.13-1.21 times those observed through the high-speed images, respectively.

\section{Experimental observations}\label{sect: Exp.}

We start with an overview of the physical phenomena observed in various experiments concerning laser-induced cavitation bubbles initiated in both oil-in-water (O/W) and water-in-oil (W/O) droplets. Moreover, we qualitatively investigate the dependencies of the overall fluid dynamics on the standoff parameter (or the eccentricity).

\subsection{Bubble initiation in an oil-in-water (O/W) droplet}\label{sect: oil-in-water}

\begin{figure}
	\centering\includegraphics[width=10cm]{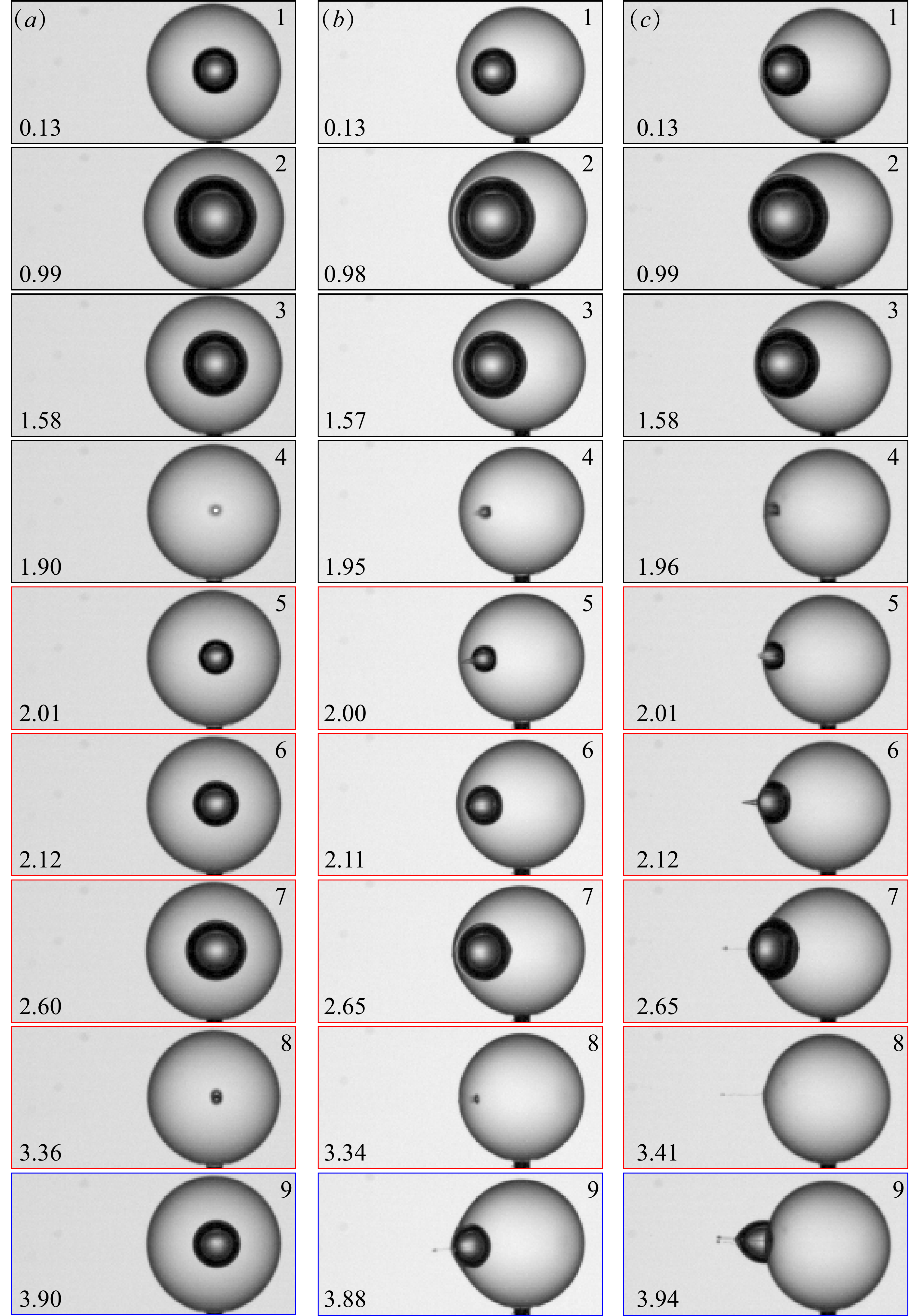}
	\caption{Three representative experiments in which bubbles were initiated within O/W droplets ($\alpha=1.093$). $(a)$ The bubble is initiated at the center of the droplet ($e = 0$, $R_{d, min } = 0.99$ mm, $R_{b, max } = 0.56$ mm, $\xi = 0.57$, $\gamma = 1.77$), maintaining a spherical shape throughout the oscillation process. $(b)$ The bubble is initiated off-center of the droplet ($e = 0.39$ mm, $R_{d, min } = 0.93$ mm, $R_{b, max } = 0.56$ mm, $\xi = 0.61$, $\gamma = 0.96$), and during the rebound stage, a liquid jet forms. $(c)$ The bubble is initiated off-center of the droplet ($e = 0.63$ mm, $R_{d, min } = 0.95$ mm, $R_{b, max } = 0.56$ mm, $\xi = 0.60$, $\gamma = 0.57$). After reaching its minimum volume, the bubble produces a rapid `needle-like' jet, piercing through the droplet surface. During the third cycle, the main body of the bubble escapes the droplet and enters the water bulk. Nondimensional times are indicated in the lower-left corners of each frame. The time scale for nondimensionalization ($R_{b, max}\sqrt{\rho_1/P_\infty}$) in these cases is 54.23 \textmu s. The horizontal width of each frame is 4.2 mm. For reader's convenience, black, red, and blue boxes represent the first, second, and third cycles of bubble oscillation, respectively.}\label{Fig:oil-in-water}
\end{figure}

Figure \ref{Fig:oil-in-water} illustrates three representative experiments in which bubbles were initiated within O/W droplets ($\alpha=1.093$). The bubble-to-droplet size ratios of the three experiments are approximately the same ($\xi \approx$ 0.6), and the bubble dynamics behavior exhibits significant variations with variations of the standoff parameter (or the eccentricity). In the first experiment shown in panel (a), the cavitation bubble originates at the center of the oil droplet, and frame 1 shows the expansion of the bubble. Frames 2, 7, and 9 capture the moments when the bubble reaches its maximum size during the first, second, and third oscillation cycles, respectively. Furthermore, frames 4 and 8 depict the instances when the bubble reaches its minimum volume at the end of the first and second cycles, respectively. Owing to the initial spherical symmetry condition, both the bubble and droplet surfaces maintain nearly spherical shapes throughout the entire process.

In the second experiment shown in Figure \ref{Fig:oil-in-water}(b), the cavitation bubble nucleates off-center and in proximity to the left interface of the droplet ($\gamma = 0.96$). Throughout the first oscillation cycle (frames 1-4), the bubble maintains a nearly spherical shape.  Notably, a high-speed liquid jet forms around the end of the first cycle of the bubble (frame 4). However, this jet fails to penetrate the droplet surface (frames 5-6). Subsequently, during the rebound stage of the third cycle, we observe the second jet that forms around the end of the second cycle finally penetrates the droplet surface (frame 9). This effectively transports oil droplets into the surrounding water bulk. The black residual produced after the jet penetration in frame 9 might be a mixture of bubble gas and oil, with an approximate diameter of 50 \textmu m (with an uncertainty of 25 \textmu m), which is about 0.03 times the droplet size.

In the third experiment shown in Figure \ref{Fig:oil-in-water}(c), the cavitation bubble is initiated closer to the left surface of the droplet ($\gamma = 0.57$). The left side of the bubble surface appears slightly flattened due to the interaction with the left surface of the droplet (frames 1-2), resulting from the higher inertia of water. Subsequently, after reaching its minimum volume, the bubble generates a rapid `needle-like jet, which pierces through the droplet surface (frames 5-6). At the same time, the bubble migrates towards the left side. During the third cycle, the main body of the bubble separates from the droplet and enters the water bulk (frame 9). As expected, this scenario promotes stronger fluid mixing compared to the second experiment.

\subsection{Bubble initiation in a water-in-oil (W/O) droplet}\label{sect: water-in-oil}

\begin{figure}
	\centering\includegraphics[width=10cm]{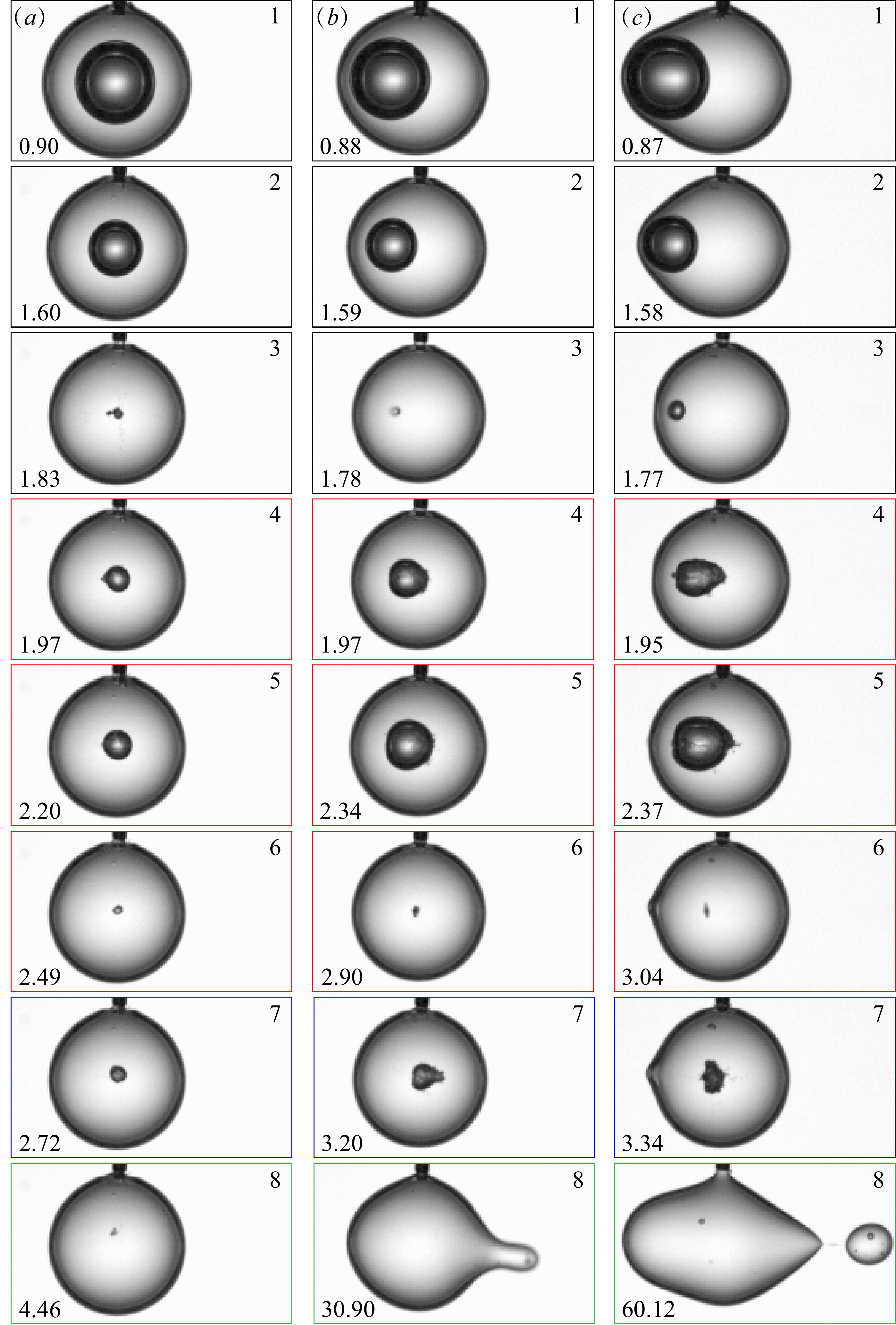}
	\caption{Three representative experiments in which bubbles were initiated in W/O droplets ($\alpha=0.915$).  $(a)$ the bubble originates at the droplet's center ($e = 0$, $R_{d, min} = 1.14$ mm, $R_{b, max } = 0.78$ mm, $\xi = 0.69$, $\gamma = 1.46$). $(b)$ The bubble is initiated off-center from the droplet's center ($e = 0.47$ mm, $R_{d, min } = 1.12$ mm, $R_{b, max } = 0.77$ mm, $\xi = 0.69$, $\gamma = 0.84$), leading to the formation of a weak jet directed towards the droplet's center. $(c)$ The bubble is initiated off-center from the droplet ($e = 0.80$ mm, $R_{d, min } = 1.14$ mm, $R_{b, max } = 0.77$ mm, $\xi = 0.68$, $\gamma = 0.44$), resulting in a tiny water-hump-like protrusion on the left side of the droplet. The residual bubble vortex continues moving to the right, causing the pinch-off of the droplet.  Nondimensional times are indicated in the lower-left corners of each frame. The time scales for nondimensionalization ($R_{b, max}\sqrt{\rho_1/P_\infty}$) are 78.99, 77.97, and 77.97 \textmu s, respectively. The horizontal width of each frame is 4.6 mm. For reader's convenience, black, red, and blue boxes represent the first, second, and third cycles of bubble oscillation, respectively, while green boxes represent the evolution of the droplet after multiple bubble cycles.}\label{Fig:water-in-oil}
\end{figure}

In this section, we present three representative experiments in which bubbles were initiated within W/O droplets ($\alpha=0.915$). In the first case, shown in Figure \ref{Fig:water-in-oil} $(a)$, the cavitation bubble originates at the center of the water droplet and oscillates in a spherical shape. However, the rebound bubble exhibits slight unstable features (frame 4) compared to that in Figure \ref{Fig:oil-in-water} $(a)$. This discrepancy arises due to the large viscosity of the oil medium, which acts as a stabilizing factor \citep{Prosperetti1977,zeng-drop} for bubble oscillations and is 50 times greater than that of water. A similar phenomena can be found in \citet{Kannan2020}, in which the dependence of the bubble dynamics on viscosity of the flow field (single-phase) was experimentally studied. Back to Figure \ref{Fig:water-in-oil} $(a)$, the nondimensional oscillation period of the bubble decreases in comparison to that in Figure \ref{Fig:oil-in-water} $(a)$. Further discussion on the influence of droplet confinement on the collapse time of the bubble will be given in \S\ \ref{sect: Rayleigh collapse time}.

In the second case, as depicted in Figure \ref{Fig:water-in-oil} $(b)$, the bubble is generated off-center within the droplet. During the rebound phase (frame 4), a weak jet emerges, directed towards the center of the droplet. Subsequently, the vortex ring bubble undergoes a migratory path from the left side to the right side of the droplet (frames 5-8). The bubble pushes the right side of the droplet surface, inducing a noticeable bulge. Nonetheless, no pinch-off of the droplet can be observed in this case. Under the influence of surface tension, the droplet progressively regains its spherical shape (not shown here).

In the third case, illustrated in Figure \ref{Fig:water-in-oil} $(c)$, the bubble forms in closer proximity to the droplet surface, resulting in a more pronounced bubble-droplet interaction. Consequently, a stronger jet forms and the bubble obtains a faster migration velocity, resulting in a breakup of the protrusion on the right side of the droplet, as evidenced in frame 8. This phenomenon could potentially constitute a secondary mechanism for fluid mixing within the current W/O system. Additionally, we notice a tiny water-hump-like protrusion on the left side of the droplet as the bubble migrates away (frames 6-7). This phenomenon is much more pronounced when the water-oil interface is initially flat \citep{Han2022}. Even for a much smaller standoff parameter case in the present bubble-droplet system, the protrusion on the left side of the droplet hardly develops to a large size and consequently no pinch-off can be found (refer to Appendix \ref{sect:appen2}).  The significant difference between the present system from \citet{Han2022} is the curvature of the water-oil interface. The dependence of the bubble-droplet interaction on the curvature parameter $\xi$ will be presented in \S\ \ref{sect: xi}.

In our current experimental setup, we have noticed that the phenomenon of jet piercing through the bubble surface is typically most prominent during the rebound phase. This observation closely resembles what has previously been characterized as the `weak-jet' phenomenon, as documented in \citet{Supponen2016}. By applying the concept of the anisotropy parameter $\zeta=0.195 \gamma^{-2}(\rho_{1}-\rho_{2})(\rho_{1}+\rho_{2})^{-1}$, as introduced by \citet{Supponen2016}, for a rough estimation of jet characteristics, we calculated $\zeta$ values of 0.012 and 0.045 for the second and third experiments in Figure \ref{Fig:water-in-oil}, respectively. While these $\zeta$ values place them within the `intermediate-jet' category, our observations more closely align with the `weak-jet' phenomenon. This discrepancy can be primarily attributed to the presence of non-flat boundaries within our experimental system, whereas the anisotropy parameter was originally derived under the assumption of flat boundaries.

\section{Spherical bubble dynamics}\label{sect: spherical}

\subsection{Comparison of experiments with theoretical results}\label{sect: exp-theory}

In this section, we discuss the dynamics of a spherical bubble located at the center of a spherical droplet. First, we try to reproduce the spherical bubble dynamics in the experiments using the theoretical model given in \S\  \ref{sec:method1}. In order to facilitate a meaningful comparison with the experimental results, it is imperative to define appropriate initial conditions for the calculations, namely $R_0$ and $\varepsilon$. We use a simplified method \citep{zeng-drop,saade2021,ZengQ2022} to initialize the bubble, setting it as an initially stationary high-pressure gas bubble that corresponds to the moment when the bubble becomes visible in the experiment. We need to adjust only the strength parameter $\varepsilon$ to match the experimental data, while the initial bubble radius is calculated using the principle of energy conservation \citep{Klaseboer04,Han2022}. A satisfactory result can be obtained if $\varepsilon$  is set as 2000. More detailed discussion on the sensitivity of the results on $\varepsilon$ can be found in Appendix \ref{sect:appen}. 

\begin{figure}
	\centering\includegraphics[width=12cm]{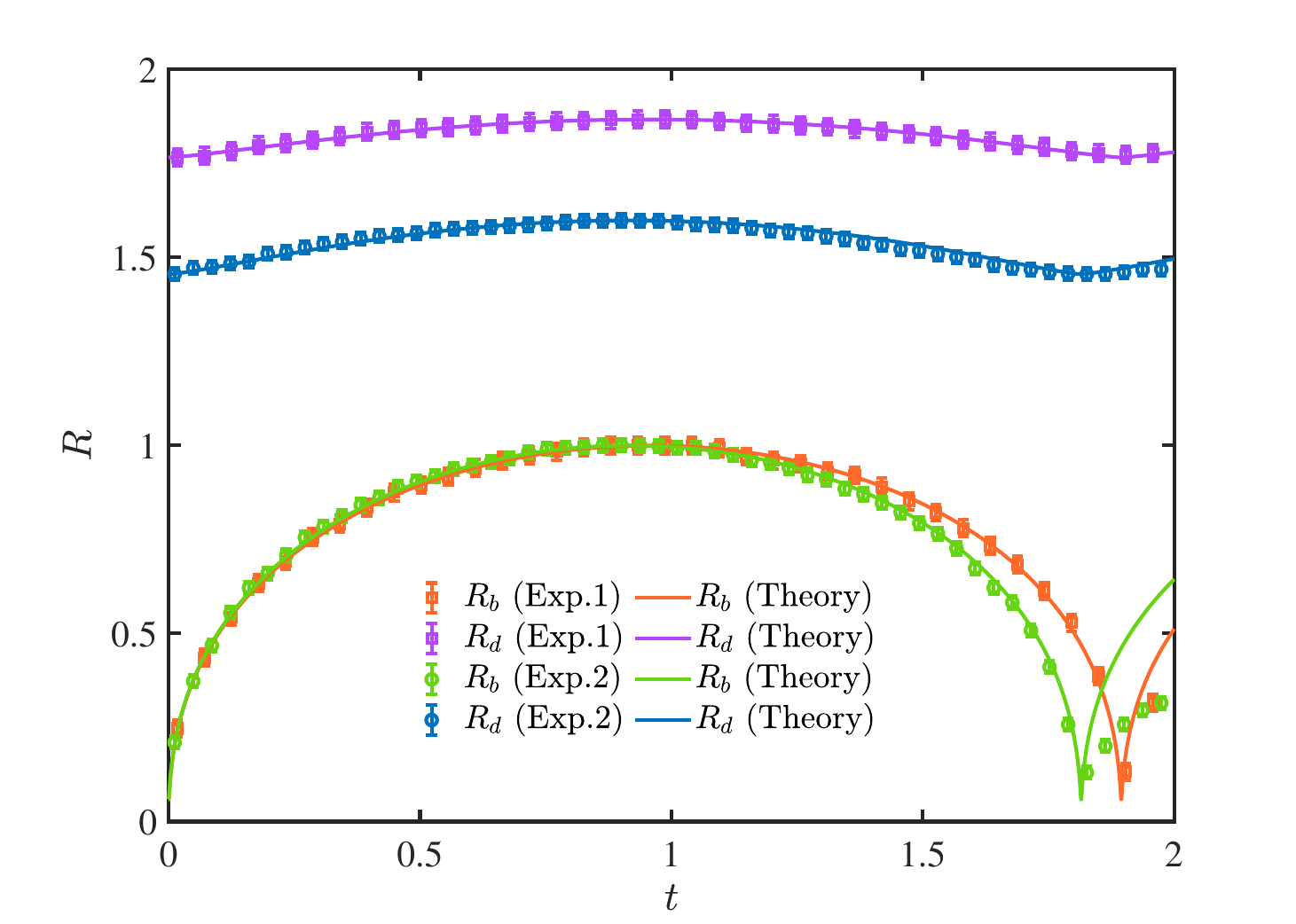}
	\caption{Time evolutions of bubble and droplet radii obtained from our experiments (denoted by the circles and rectangles with error bar) and the theoretical model (denoted by the solid lines). In the first experiment, the bubble was initiated at the center of an O/W droplet, with the following parameters: $R_{d, min } = 0.99$ mm, $R_{b, max } = 0.56$ mm, $\xi = 0.57$, $\gamma = 1.77$. In the second experiment, the bubble was initiated at the center of a W/O droplet, with the following parameters: $R_{d, min } = 1.14$ mm, $R_{b, max } = 0.78$ mm, $\xi = 0.69$, $\gamma = 1.46$. The time scales for nondimensionalization ($R_{b, max}\sqrt{\rho_1/P_\infty}$) of the two experiments are 54.23 and 78.99 \textmu s, respectively. The nondimensional initial gas pressure and radius of the bubble in computations are set as: $\varepsilon=2000$ and $R_0=0.0591$.
	}\label{Fig:spherical-comparison}
\end{figure}

Figure \ref{Fig:spherical-comparison} displays the time evolutions of bubble and droplet radii for the two experiments shown in Figure \ref{Fig:oil-in-water} $(a)$ and Figure \ref{Fig:water-in-oil} $(a)$. As can be seen, our theoretical model accurately predicts the bubble and droplet dynamics. Note that the nondimensional bubble oscillation period deviates a lot between the two experiments. This is mainly attributed to the difference in the density ratio $\alpha$. We will discuss the influences of the density ratio $\alpha$ and size ratio $\xi$ on the bubble collapse time in \S\ \ref{sect: Rayleigh collapse time}. Although the numerical results obtained from the boundary integral (BI) method are not shown here (for the sake of image clarity), we want to emphasize that the difference between the results obtained from BI simulations and the theoretical model is indistinguishable. 

\subsection{Modified Rayleigh collapse time}\label{sect: Rayleigh collapse time}

\begin{figure}
	\centering\includegraphics[width=14cm]{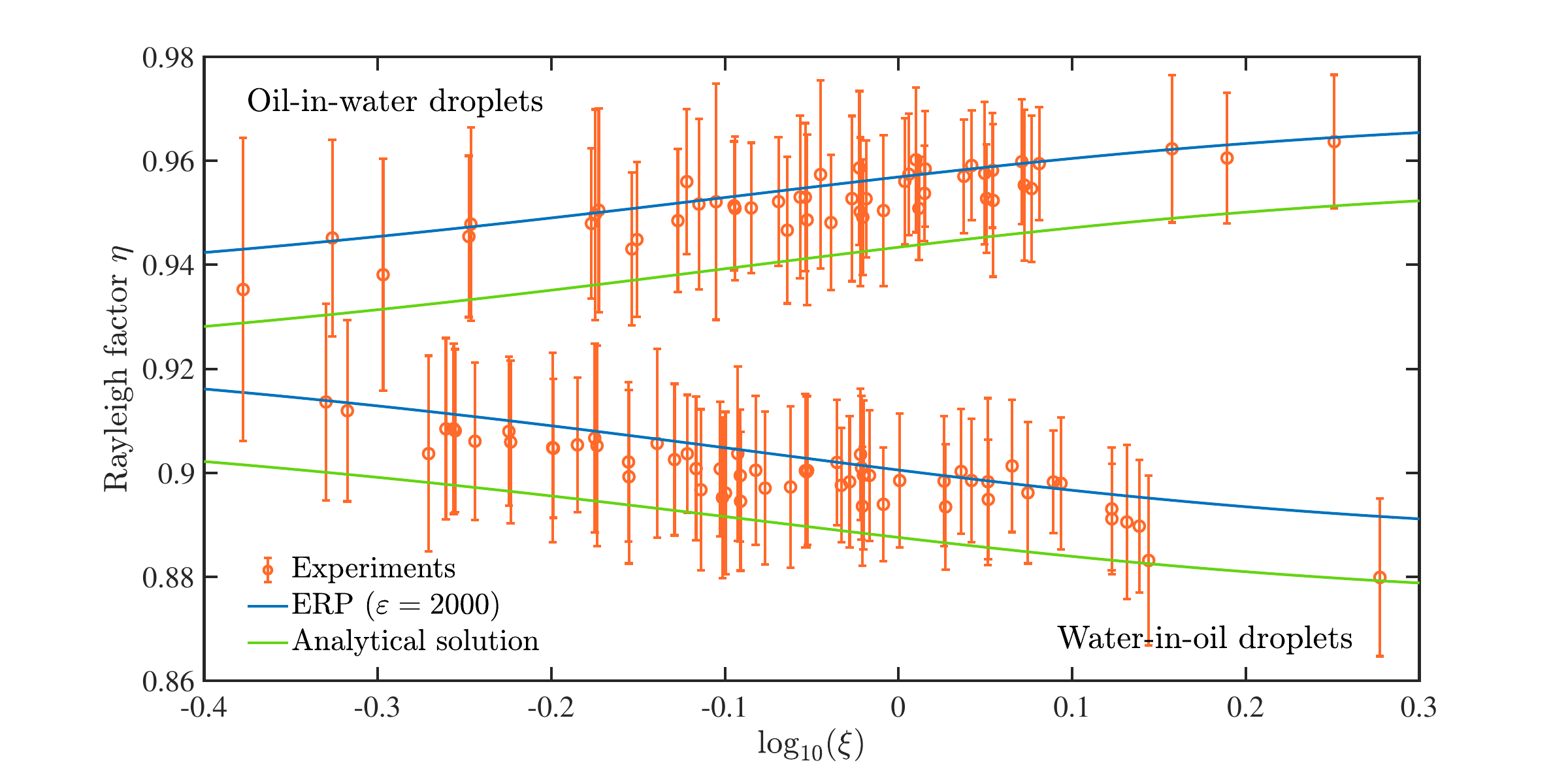}
	\caption{Variation of the Rayleigh factor $\eta$ over the size ratio $\xi$. The upper and lower parts of the figure represent two different situations of bubble generation within O/W droplets ($\alpha = 1.093$) and W/O droplets ($\alpha = 0.915$), respectively. The orange circles represent the experimental data, the blue solid lines represent the results obtained from ERP equation ($\varepsilon = 2000$), and the green solid lines denote the analytical solution (calculated from Equation \ref{Equation:factor}).}\label{Fig:eta-exp}
\end{figure}

\begin{figure}
	\centering\includegraphics[width=13cm]{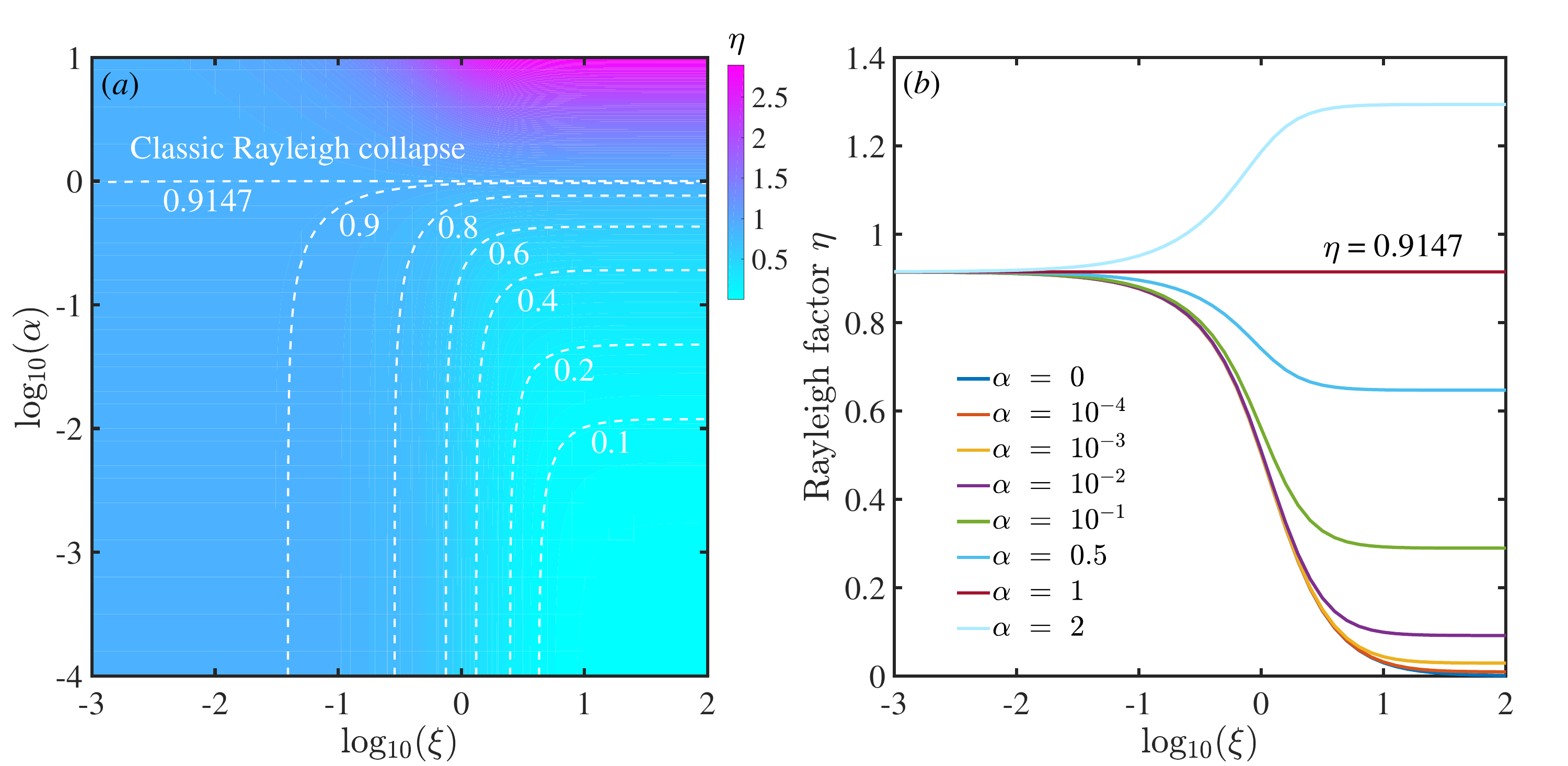}
	\caption{Variation of the Rayleigh factor $\eta$ in the $\xi-\alpha$ space. In panel (a), the white dashed lines denote isolines of $\eta = $ 0.1, 0.2, 0.4, 0.6, 0.8, 0.9, and 0.9147 (the classic Rayleigh factor), respectively. In panel (b), the case where $\alpha=0$ corresponds to the result obtained by \citet{Obreschkow}. They considered a constant pressure boundary condition on the droplet surface but neglected the influence of air flow.}\label{Fig:eta}
\end{figure}

 \citet{Rayleigh} first derived the analytical expression for the collapse time of a vacuum bubble in an infinite medium, namely, $T_c=\eta R_{b,  max}\sqrt{{\rho_1}/{\Delta P}}$, where the Rayleigh factor $\eta$ is approximately 0.9147.  \citet{Obreschkow} revealed a remarkable reduction in the collapse time for a bubble collapsing inside a droplet surrounded by air. In this study, we derive a modified Rayleigh collapse time for bubbles within a droplet, which is characterized by the density ratio $\alpha$ and bubble-to-droplet size ratio $\xi$. 
 
To start, we compare the experimental results with the theoretical predictions. In our experimental trials, we can adjust the bubble-to-droplet size ratio in the range of $0.4 < \xi < 2$. The Rayleigh collapse time in our experiments, denoted as $T_c$, is defined as half of the first oscillation period of the bubble. The Rayleigh factor $\eta$ is calculated using the formula $T_c/\left( R_{b, max} \sqrt{{\rho_1}/{P_{\infty}}}\right)$. For the purpose of comparison, we analyze the experimental data alongside both the analytical estimation and the results obtained from the extended Rayleigh-Plesset equation (ERP). These computations encompass scenarios in which bubbles are initiated within both W/O and O/W droplets, and the results are depicted in Figure \ref{Fig:eta-exp}. Similar to the results illustrated in Figure \ref{Fig:spherical-comparison}, the strength parameter is consistently set at $\varepsilon=2000$ for calculations utilizing ERP. In cases where bubble initiation occurs within W/O droplets ($\alpha=0.915$), it is evident that the Rayleigh factor $\eta$ exhibits a decreasing trend with increasing $\xi$. It is important to note that the analytical estimation provides a lower limit due to the assumption that the bubble's interior is a vacuum. Notably, the outcomes obtained through the ERP align well with the experimental data. Conversely, for bubbles initiated within O/W droplets ($\alpha=1.093$), there is a noticeable increase in the Rayleigh factor $\eta$ as $\xi$ increases.

Next, we evaluate the modified Rayleigh factor over an extended parameter space, specifically $10^{-4}\le\alpha\le10$ and $10^{-3}\le \xi \le10^2$. As shown in Figure \ref{Fig:eta}(a), the contour represents the value of $\eta$ in the $\xi-\alpha$ space. This plot can be interpreted in several ways. First, when $\alpha=1$, the classic Rayleigh collapse factor is obtained, as indicated by the upper horizontal line $\eta=0.9147$. Above this line ($\alpha>1$), $\eta$ increases with $\alpha$, particularly for larger values of $\xi$. This is because the heavier outer fluid (fluid 2) is more difficult to accelerate, resulting in an increase in the bubble collapse time. Notably, the influence of $\alpha$ becomes increasingly pronounced as $\xi$ rises, corresponding to a reduction in droplet size. Second, for very small size ratios $\xi<10^{-2}$, $\eta$ is almost independent of $\alpha$, implying that the outer fluid phase has little effect on the bubble if the droplet is much larger than the bubble. Third, the Rayleigh factor decreases with decreasing $\alpha$ and increasing $\xi$, corresponding to a reduction in the mass of fluid that needs to be accelerated by the bubble.

Figure \ref{Fig:eta}(b) depicts the variation of $\eta$ with respect to $\xi$ for different $\alpha$. When $\alpha<1$, we observe that $\eta$ decreases slowly in two different parameter regimes, i.e., $\xi\lesssim10^{-1}$ and $\xi\gtrsim10$, while it decreases rapidly in the range of $10^{-1}\lesssim\xi\lesssim10$. Thus, if there is a large size difference between the bubble and droplet, the bubble dynamics are mainly dominated by the inertia of one fluid. However, when the bubble and droplet have comparable size, the Rayleigh factor is more sensitive to the variation of $\xi$. Additionally, the criterion for neglecting the effect of the outer phase on bubble dynamics is worth discussing. The case of $\alpha=0$ (the lowest blue line) corresponds to the result given by \citet{Obreschkow}. For comparison, let us take the case of $\alpha=10^{-3}$ (which corresponds to air and is denoted by the yellow line). The results for the $\alpha=0$ and $\alpha=10^{-3}$ cases are almost identical when $\xi\lesssim1$, and deviations can be observed only when $\xi>10$. This suggests that the effect of air can be neglected for small bubbles in a large droplet.

\section{Nonspherical bubble dynamics}\label{sect: nonspherical}

In this section, we delve deeper into the dynamics of nonspherical bubbles within droplets. Initially, we validate our boundary integral (BI) model through meticulous one-to-one comparisons between experimental data and numerical simulations. Subsequently, we aim to elucidate the influence of key parameters governing bubble jetting behavior.

\subsection{Comparison between experimental and BI simulation results}\label{sect: com_jet}

\begin{figure}
	\centering\includegraphics[width=12cm]{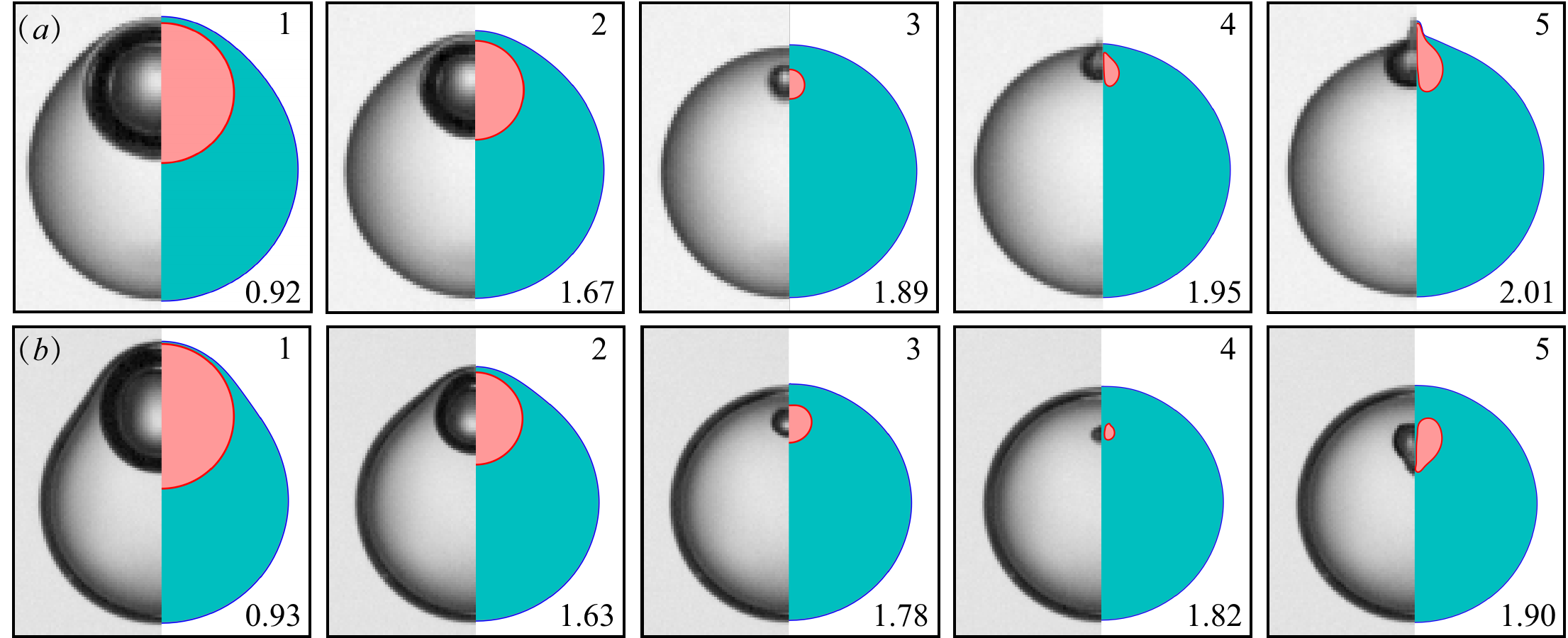}
	\caption{Comparison between experimental observations (left-hand half of each frame) and BI simulations (right-hand half of each frame) for a jetting bubble inside a droplet. $(a)$ O/W droplet with the following parameters: $R_r=0.96\ \rm mm,\ \it R_z=\rm 0.95\  mm$, $e = 0.61$ mm, $R_{b, max} = 0.53$ mm, and $\alpha=1.093$. The width of each frame is 1.10 mm, and the time scale for nondimensionalization is 51.33 \textmu s. $(b)$ W/O droplet with the following parameters: $R_r=1.25\ \rm mm,\ \it R_z=\rm 1.20\ mm$, $e = 0.92$ mm, $R_{b, max} = 0.75$ mm, and $\alpha=0.915$. The width of each frame is 1.50 mm, and the time scale for nondimensionalization is 75.95 \textmu s. Nondimensional times are indicated in the lower right corner. The times of the BI simulation for frames 4-5 in panel (a) are 1.93 and 1.99, respectively, while the times for other frames match those of the experiment. Due to light refraction, the actual sizes of the bubbles in these two experiments are 0.97 and 1.15 times the sizes observed in the high-speed images, respectively. For the convenience of comparison, all the experimental images are rotated $90^\circ$ clockwise.}\label{Fig:exp-jet}
\end{figure}

In Figure \ref{Fig:exp-jet}, we present a comparison between experimental observations and BI simulations regarding bubble and droplet profiles. Each frame is divided into two halves: the left side displays the experimental observations, while the right side shows the simulation results. The bubble and droplet profiles are depicted using solid red and blue lines, respectively. In Figure \ref{Fig:exp-jet} (a), a bubble is generated inside an O/W droplet, characterized by the following parameters: $R_r=0.96$ mm, $R_z$ = 0.95 mm, $R_{b, max}=0.53$ mm, $e = 0.61$ mm, $\gamma=0.64$, and $\alpha=1.093$. In frame 1, as the bubble grows to its maximum size in an almost spherical shape, no interfacial instability is observed on the droplet surface. During the subsequent collapse stages (frames 2-3), the bubble retains its nearly spherical shape due to the minimal density difference between the two liquids. Simultaneously, the droplet gradually regains its spherical form. In the final collapse phase, between frames 3 and 4, the bubble generates a thin liquid jet that propels towards the upper interface of the droplet. By frame 5, the jet tip ascends and ultimately impacts the droplet interface, transporting oil into the surrounding water bulk. It's worth noting that the times of the BI simulation for frames 4-5 in panel (a) are 1.93 and 1.99, respectively, while the times for frames 1-3 align with those of the experimental observations. The slight deviation between the simulation and experiment during the rebounding stage of the bubble can be attributed to the omission of energy loss in the current BI model. Nonetheless, our numerical model effectively captures the fundamental physics underlying the interaction between the bubble and droplet, particularly the critical process of the jet impacting the droplet surface. Further quantitative comparisons of jet velocity and maximum impact velocity on the droplet surface will be presented in \S\ \ref{sect:discussion1}.

In Figure \ref{Fig:exp-jet} (b), we examine the generation of a bubble in close proximity to the interface of a W/O droplet, characterized by the following parameters: $R_r=1.25$ mm, $R_z$ = 1.20 mm, $R_{b, max}=0.75$ mm, $e = 0.92$ mm, $\gamma=0.37$, and $\alpha=0.915$. As the bubble grows to its maximum size, a significant curvature protrusion emerges on the upper surface of the water droplet (frame 1). This phenomenon arises because the fluid's inertia around the upper surface of the bubble is smaller compared to other directions. Notably, we observe a thin water film separating the bubble from the oil, which poses a numerical challenge. To ensure precise simulation, the element size on both surfaces must be smaller than the liquid layer's thickness. To tackle this issue, we employed 500 elements on both the bubble and droplet surfaces, ensuring accuracy and stability in the simulation. Throughout most of the collapse stage (frames 2-3), the bubble maintains an approximately spherical shape. Subsequently, a jet forms toward the center of the droplet at approximately the moment when the bubble reaches its minimum volume (frame 4). Thereafter, the toroidal bubble rapidly rebounds and migrates downwards (frame 5). In this case, due to light refraction, the actual size of the bubble is about 1.15 times the size observed in the high-speed images. As evident, our BI model also reproduces the primary features of this experiment.

\subsection{Dependence of bubble dynamics on the droplet curvature $\xi$}\label{sect: xi}

Previous research had examined the interaction between a cavitation bubble and a planar fluid-fluid interface  \citep{Chahine1980,Klaseboer-CM2004,Orthaber,Han2022}.  For the present bubble-droplet system, the curvature of the droplet surface is an important parameter. The impact of water-air interface curvature on bubble dynamics has been emphasized \citep{Obreschkow}, and therefore we aim to examine the relationship between bubble dynamics and the  bubble-to-droplet size ratio $\xi$ (denotes half of the droplet curvature) via a series of boundary integral simulations. The simulation results discussed below encompass a broad parameter range: $10^{-3}\le\xi\le1.5$, with $\gamma=0.6$ being the fixed standoff parameter. We will further explore the effect of $\gamma$ in \S\ \ref{sect: parameter space}.

Figure \ref{Fig:xi}(a) shows the overall interaction between the bubble and droplet for a typical value of $\xi=0.3$ and two different density ratios, $\alpha=0.8$ (left) and $\alpha=1.2$ (right). When the density of the host fluid is smaller than the droplet density ($\alpha=0.8$), the bubble is repealed by the nearby droplet surface, resulting in a high-speed liquid jet directed away from the interface during the collapse phase. Conversely, when the density of the host fluid is larger than the droplet ($\alpha=1.2$), the jet moves towards the droplet surface. Notably, the latter scenario, with a larger $\alpha$, contributes more to fluid mixing from the perspective of jet penetration. To evaluate the asymmetry of the collapse, we introduce the concept of Kelvin impulse (expressed in nondimensional form, with the density term omitted here):

\begin{figure}
	\centering\includegraphics[width=14cm]{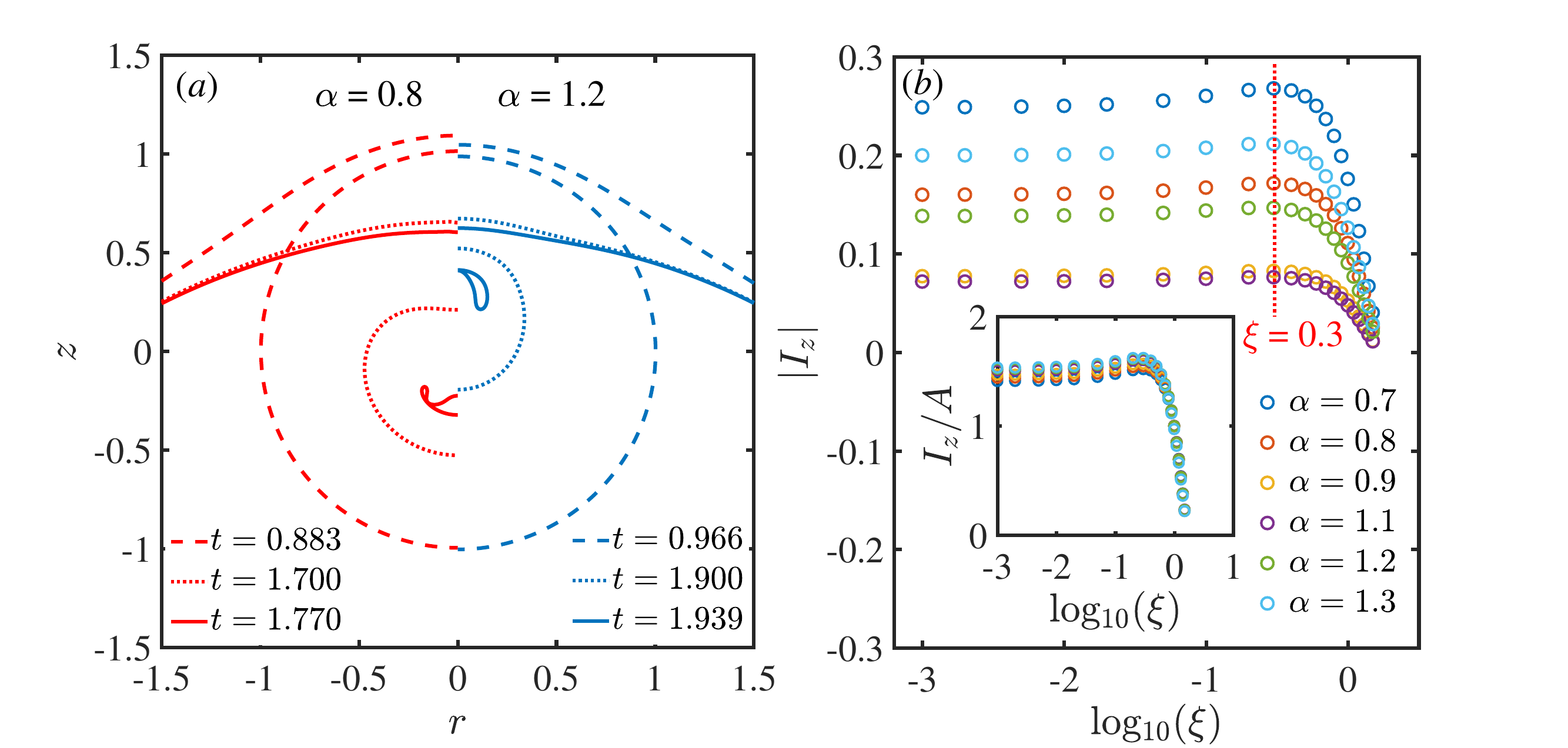}
	\caption{Quantitative study on nonspherical bubble dynamics inside a droplet for different $\alpha$ and $\xi$. (a) Bubble collapse patterns inside a droplet with $\alpha=0.8$ (on the left-half) and $\alpha=1.2$ (on the right-half). The standoff parameter is fixed at $\gamma=0.6$ and the size ratio is $\xi=0.3$. (b) Variations of the Kelvin impulse of the bubble $|I_z|$ at the jet impact moment versus $\xi$. Note that $|I_z|$ or $I_z/A$  is maximized at  $\xi=0.3$, independent of $\alpha$. The inset shows a normalized $I_z$ by the Atwood number $A=(\alpha-1)/(\alpha+1)$.}\label{Fig:xi}
\end{figure}

\begin{equation}
\textbf{\textit{I}}=\int_{S_{ b}} {\varphi_1 \textbf{\textit{n}}\rm d\it S}. \label{Equation:Kelvin}
\end{equation}
 
Since we limited ourselves to the axisymmetric configuration, only the vertical component of the Kelvin impulse $I_z$ is considered in this context. Figure \ref{Fig:xi}(b) shows the variation of $|I_z|$ over $\xi$ for different $\alpha$. As $\alpha$ approaches 1, which indicates a reduction in the asymmetry of the bubble collapse, the value of $|I_z|$ diminishes. Previous studies \citep{Blake2015,Supponen2016,Han2022} have shown that the nondimensional Kelvin impulse of a bubble near a flat fluid-fluid interface scales with the Atwood number $A=(\alpha-1)/(\alpha+1)$. This can be expressed as follows:
\begin{equation}
I_z\propto \frac{\alpha-1}{\alpha+1}. \label{Equation:At}
\end{equation}

As shown in the inset of Figure \ref{Fig:xi}(b), we are surprised that the data nearly collapse together when we normalized $I_z$ with the Atwood number. This suggests that the scaling of Equation (\ref{Equation:At}) still holds well for the current curved fluid-fluid interface in a large parameter space.

Next, we examine how bubble dynamics are affected by $\xi$.  As $\xi$ increases from a very small value (where the fluid-fluid interface is nearly flat), $|I_z|$ initially increases and then decreases.  Interestingly, $|I_z|$ or $I_z/A$ is maximized at $\xi \approx 0.3$, independent of $\alpha$. Compared to a flat fluid-fluid interface, a curved interface is overall closer to the bubble, resulting in stronger bubble-droplet interaction.  Therefore, the value of $|I_z|$ increases with increasing $\xi$ at first. When $\xi$ is very small, the droplet is relatively large, and the lower surface of the droplet is far from the bubble, while the upper surface of the droplet is primarily responsible for the asymmetrical motion of the bubble. However, when $\xi$ exceeds 0.3, the droplet size gradually becomes comparable to the bubble, and the lower surface of the droplet also influences bubble dynamics. Consequently, $|I_z|$ decreases with $\xi$ when $\xi$ is greater than 0.3. This critical value of $\xi \approx 0.3$ applies to a wide range of parameters: $0.5 \leq \alpha \leq 1.6$ and $\gamma \leq 1.1$.

\subsection{Bubble dynamics in the $\alpha-\gamma$ parameter space}\label{sect: parameter space}
In the current system, the interaction between bubbles and droplets is primarily governed by three key parameters: $\xi$, $\alpha$, and $\gamma$. While we discussed the influence of $\xi$ in the previous section, our attention now shifts to exploring the dynamics of bubbles within the $\alpha-\gamma$ parameter space. In this section, our focus narrows to scenarios with small standoff parameters, where $\gamma\leq1$. This choice is motivated by the fact that for large standoff parameters ($\gamma\gtrsim0.9$), as will be discussed in \S\ \ref{sect:discussion1}, the jet cannot effectively penetrate water-oil interface. We will delve further into the critical standoff parameter required for successful jet penetration of the droplet surface. 

\begin{figure}
	\centering\includegraphics[width=12cm]{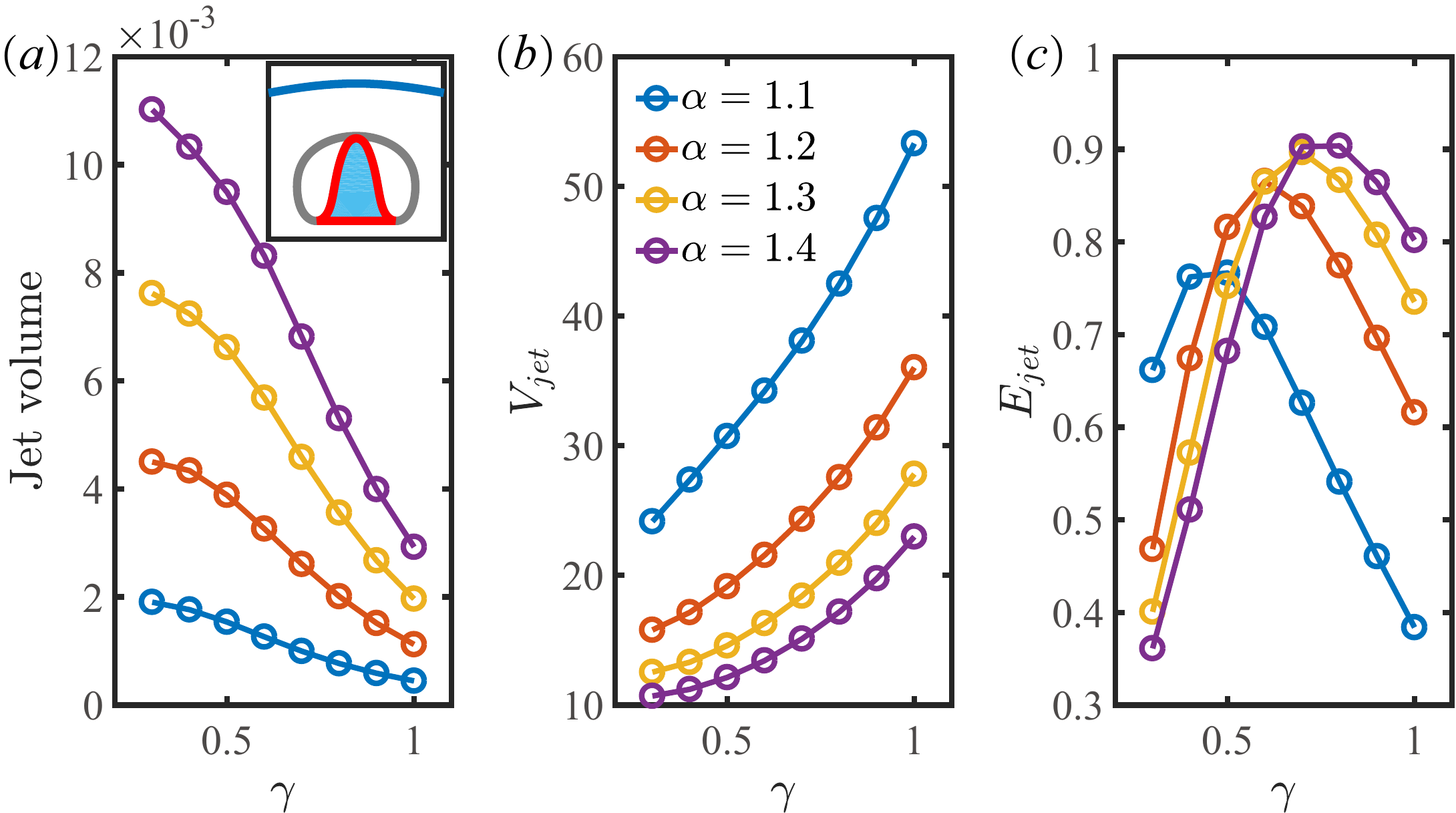}
	\caption{Dependence of bubble dynamics on $\alpha$ and $\gamma$ at a fixed $\xi=0.6$. (a) Bubble jet velocity, (b) jet volume and (c) kinetic energy of the jet. The upper inset of panel (a) shows a sketch of the bubble jet. All the panels share the same legend with panel (b).}\label{Fig:xi0.6}
\end{figure}

Figure \ref{Fig:xi0.6} illustrates the relationship between bubble jet volume, jet velocity, and kinetic energy concerning $\gamma$ and $\alpha$ while maintaining $\xi$  at a typical value of 0.6. These parameters provide valuable insights into the mass transport capabilities of the bubble. Figure \ref{Fig:xi0.6} (a) depicts the variation of jet volume with respect to $\gamma$, accompanied by a schematic representation of the jet volume in the upper inset. Notably, increasing $\alpha$ results in a proportional increase in jet volume and inertia. This indicates that a higher $\alpha$ value can yield a larger jet. Moving on to Figure \ref{Fig:xi0.6} (b), we examine the behavior of jet velocity ($V_{jet}$), defined as the velocity of the jet tip just prior to impact. Here, we observe that $V_{jet}$ decreases as $\alpha$  increases but increases with $\gamma$. This behavior is attributed to the weakening of jet acceleration as the jet's volume and mass expand.

Following \citet{Pearson}, \citet{LiS2020} and \citet{Han2022}, we introduce the kinetic energy associated with the liquid jet:
 \begin{equation}
 E_{jet}=\frac{1}{2}\  \rho_1 \int_{S_{jet}} {\varphi_1\frac{\partial \varphi_1}{\partial n}\rm d\it S}, \label{Equation:Ejet}
 \end{equation}
where $S_{jet}$  represents a closed surface delineating the confinement of the liquid jet within the bubble. This surface is depicted by the red solid line in the upper inset of panel (a). In Figure \ref{Fig:xi0.6}(c), we explore the dependence of $E_{jet}$  on the parameters $\gamma$ and $\alpha$. Notably, due to its incorporation of mass (or liquid volume) and velocity components, the kinetic energy, $E_{jet}$, does not exhibit a straightforward monotonic relationship with $\gamma$. However, it attains its maximum value at an optimal standoff parameter, denoted as $\gamma_o$. It is noted that $\gamma_o$ increases with the parameter $\alpha$.

\begin{figure}
	\centering\includegraphics[width=12cm]{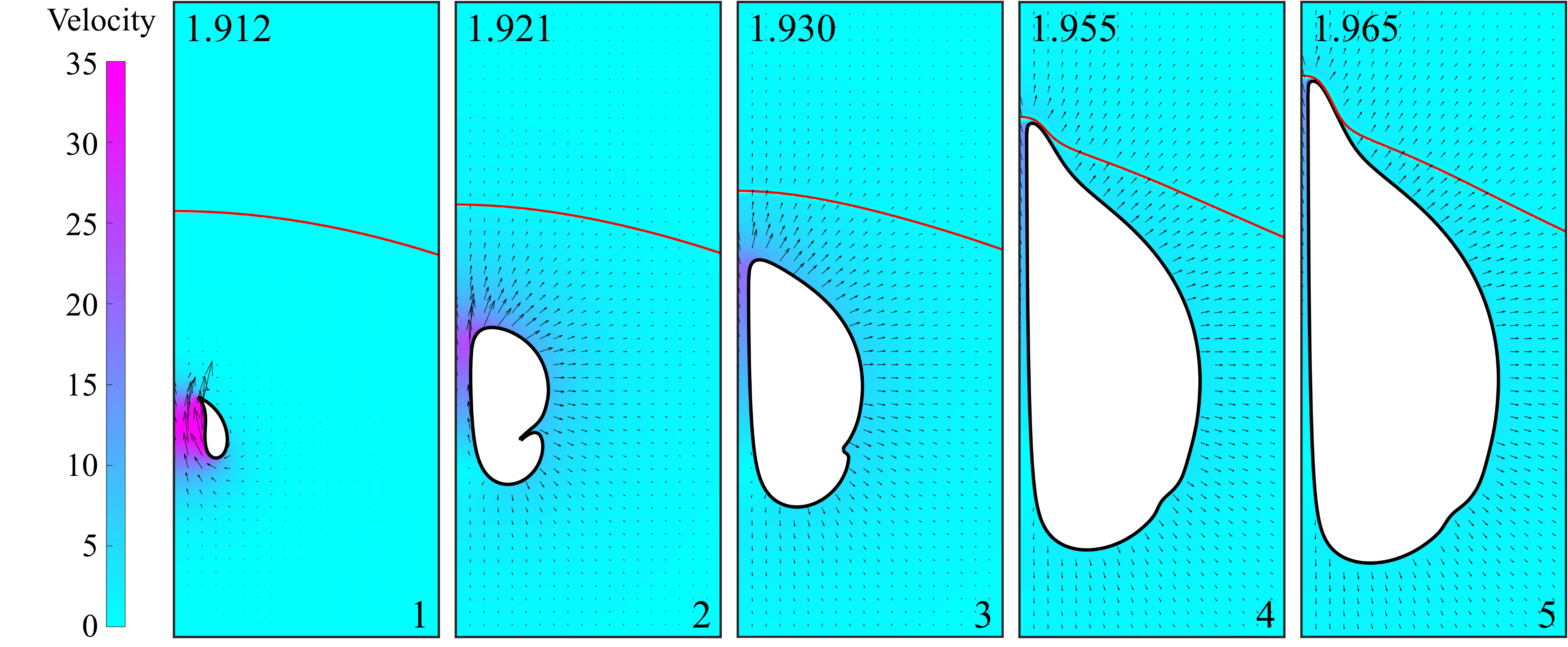}
	\caption{Evolution of a toroidal bubble and the jet impact on the droplet surface for $\xi=0.6$, $\gamma=0.6$, and $\alpha=1.1$. The contour denotes the magnitude of the flow velocity. The panels in the figure feature a unified legend on the left side. Horizontal and vertical axes are $0\le r\le 0.5$, $-0.2\le z \le 1$. The dimentionless times are marked at the upper left corners.}\label{Fig:contour1}
\end{figure}

After the jet impacts the bubble, it transforms into a toroidal shape. To model this phenomenon, we introduce a vortex ring within the bubble. Additional details regarding the numerical model are available in \citet{WangQX1996} and \citet{Zhang2015pof}.  Figure \ref{Fig:contour1} shows the evolution of the toroidal bubble and the corresponding droplet deformation under the jet impact with $\gamma=0.6$, $\xi=0.6$, and $\alpha=1.1$. The contour denotes the magnitude of the flow velocity. As shown in frame 1 ($t=1.912$), the velocity within the bubble jet is much higher than in other positions of the flow field. Subsequently, the liquid jet goes upward and pushes the droplet surface outward rapidly (frames 2-5). The jet cannot impact the droplet surface directly as the liquid layer between the bubble top and the droplet surface acts like a cushion. The bubble rebounds at this stage, with both the width and jet velocity decreasing. The velocity at the top of the droplet surface (denoted by $V_{d}$) initially increases and reaches a peak velocity as the jet tip approaches the droplet surface (frame 4). After the jet starts to enter the surrounding liquid phase, $V_{d}$ rapidly decreases (frame 5). In the numerical simulation, the liquid jet can penetrate deeply into the surrounding phase until the thickness of the liquid layer between the bubble and droplet becomes less than the grid size. It appears that the liquid jet has the potential to transport the droplet phase into the outer phase, which may result in the production of very fine droplets. Our BI model struggles to capture the fragmentation of the liquid jet that follows. More discussion on the jet penetration of the droplet surface from the experimental aspect will be provided in \S\  \ref{sect:discussion1}.

\section{More discussion}\label{sect:discussion}

\begin{figure}
	\centering\includegraphics[width=12cm]{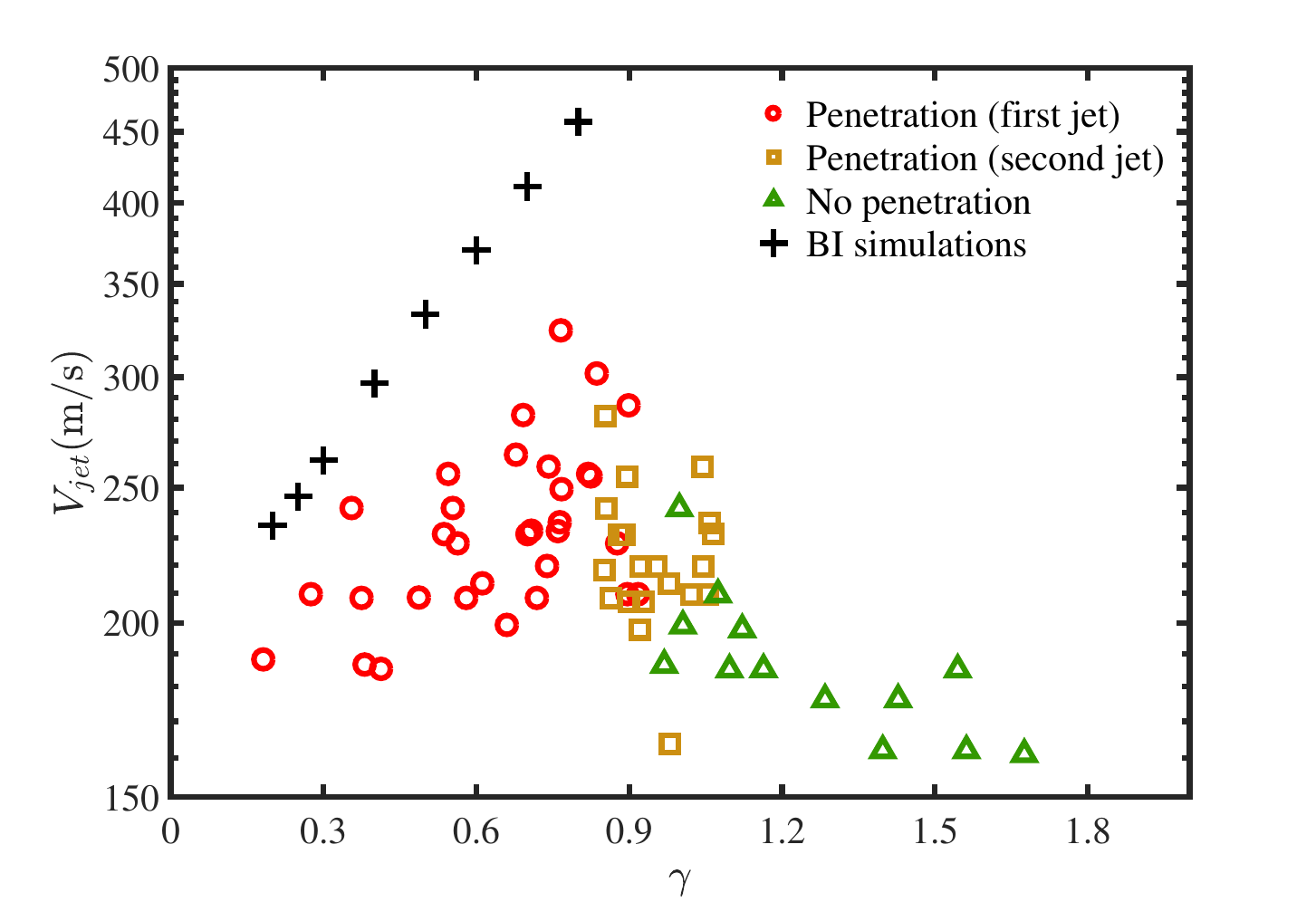}
	\caption{Variation in jet impact velocity with respect to the standoff parameter $\gamma$. The red circles and yellow rectangles signify the penetration of the droplet surface by the first and second bubble jets, respectively. Meanwhile, the green triangles indicate scenarios where the bubble jet cannot penetrate the droplet surface. The critical standoff parameter $\gamma_{c}$, which divides the first two regimes of jet behaviors, lies within the range of 0.85 to 0.91. If $\gamma \gtrsim 1.07$, the jet cannot penetrate the droplet surface during the whole lifetime. The bubble-to-droplet size ratios of these experiments range from 0.58 to 1.05. The parameters in BI simulations are set as: $\varepsilon = 2000$, $R_0=0.0591$, $\alpha=1.093$, and $\xi=0.6$.}\label{Fig:penetration}
\end{figure}

\subsection{Jet penetration}\label{sect:discussion1}
As discussed in \S\  \ref{sect: oil-in-water}, a very thin liquid jet forms around the end of the bubble collapse phase, culminating in the jet impacting the droplet surface. This impact creates a sharp protrusion on the droplet  surface, which may explain how fine droplets are produced in the O/W system. In our experiments, we notice that the liquid jet can penetrate the droplet interface at a relatively small standoff parameter. If we only consider the phenomena during the first collapse and the subsequent rebound process of the bubble, the upper bound of the standoff parameter (denoted by $\gamma_c$) for jet penetration of the droplet surface is about 0.88$\pm 0.03$, which is summarized from more than 60 experiments, as shown in Figure \ref{Fig:penetration}. Interestingly, this critical value is very close to the configuration with an initially flat fluid-fluid interface and a similar density ratio \citep{Han2022}. This implies that the jet penetration condition is insensitive to the curvature of the fluid-fluid interface in the parameter range of the present experiment ($0.58 < \xi < 1.05$). Additionally, if the standoff parameter ranges from 0.85 to 1.05, we may observe the jet penetration of the droplet surface when the bubble forms a second jet around the end of the second cycle, as denoted by the yellow rectangles in Figure \ref{Fig:penetration}. When $\gamma\gtrsim1.07$, we can hardly observe the jet penetration of the droplet surface during the whole bubble lifetime, which implies that no fluid mixing takes place.

\begin{figure}
	\centering\includegraphics[width=12cm]{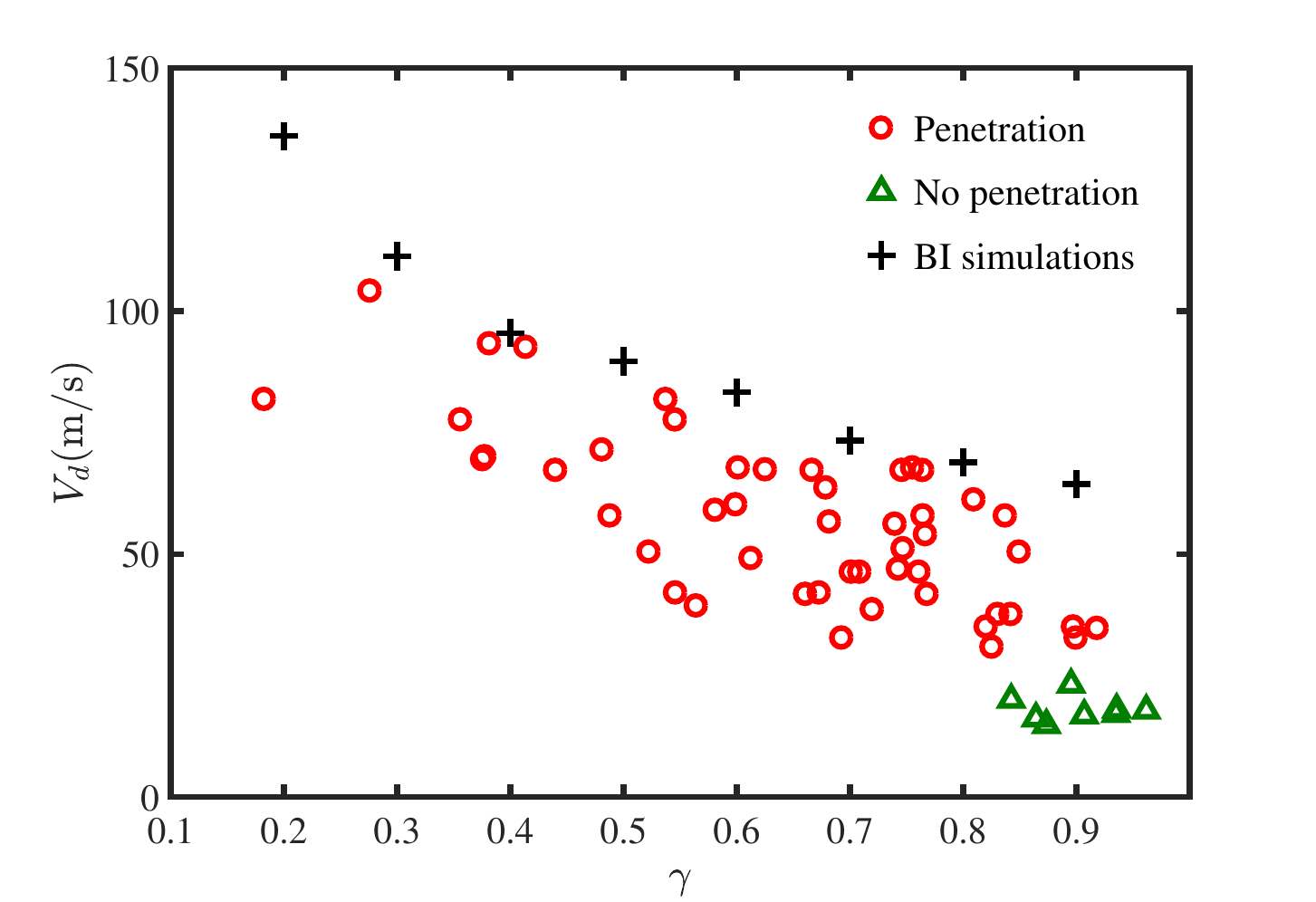}
	\caption{Dependence of the maximum velocity at the impact position of the droplet surface (denoted by $V_{d}$) on $\gamma$. 
	The red circles signify the penetration of the droplet surface by the first bubble jet that forms around the end of the first cycle. The green triangles indicate scenarios where the first bubble jet cannot penetrate the droplet surface. The minimum value of $V_{d}$ for jet penetration is about 31 m/s. The bubble-to-droplet size ratios of these experiments range from 0.58 to 1.05. The numerical results (denoted by the black cross) are also added for comparison, which align well with the upper bound of the experimental data. The parameters in BI simulations are set as: $\varepsilon = 2000$, $R_0=0.0591$, $\alpha=1.093$, and $\xi=0.6$. 
	}\label{Fig:penetration2}
\end{figure}

Due to the limited spatio-temporal resolution of high-speed recordings, we can hardly trace the jet head within the bubble interior. Hence, the jet impact velocity can be roughly evaluated from two adjacent frames before and after the jet impact moment. These measured velocities should be considered as lower bounds of the actual values. Nevertheless, the maximum value of jet velocity reaches more than 300 m/s. We also added the results obtained from BI simulations in Figure \ref{Fig:penetration}. As anticipated, the numerical results exceed the experimental data. We notice that the jet impact velocity increases with $\gamma$ in BI simulations, as already discussed and explained in \S\  \ref{sect: parameter space}. However, in the experiments, $V_{jet}$ has a tendency of increase in the range of $\gamma<0.9$ and decrease in the range of $\gamma>0.9$. This phenomenon can be explained as follows. When $\gamma>0.9$, the width and kinetic energy of the bubble jet decreases rapidly with $\gamma$. Although the jet possesses a very high velocity prior to the moment of impact, it would experience rapid deceleration after the impact. Take the $\gamma=0.6$ case for example, we find from the BI simulation that the jet velocity increases from 100 m/s to 360 m/s within 0.87 \textmu s, and then decreases rapidly after the jet impact. However, the time interval is more than 1.5 \textmu s in our experiments. It is a very challenging work for us to measure the jet velocities accurately at present. Nevertheless, our BI simulations can provide a reference for the high-speed liquid jet formation.

As inferred from previous discussion, the jet cannot impact the droplet surface directly as the liquid layer between the bubble and the droplet surface acts like a cushion. Therefore, we measure the maximum velocity at the impact position of the droplet surface (denoted by $V_{d}$), which may be a more direct way to quantify the jet penetration process that is responsible for fluid mixing. Figure \ref{Fig:penetration2} shows the dependence of $V_{d}$ on $\gamma$ and the numerical results are also added for comparison. The maximum value of $V_{d}$ reaches more than 100 m/s. Although the data have a certain degree of discreteness, the plot reveals an overall decrease of $V_{d}$ over $\gamma$. Remarkably, our BI simulation reproduces the upper bound of the experimental data quite well when $\gamma\lesssim0.8$. We note that the value of $V_d$ at $\gamma=0.9$ measured in the experiments is overestimated by our BI model. This is because the jet tip reaches the droplet surface when the bubble rebounds to the maximum size in this case, thus the impact velocity on the droplet is much influenced by the energy loss of the rebounding bubble, which is not considered in the present BI model. Finally, our experiments demonstrate that the minimum value of $V_{d}$ for jet penetration is about 31 m/s. In the work by \citet{Yamamoto2021}, a condition for droplet fragmentation was proposed, which relies on a balance between interfacial and kinetic energies, along with a comparison between dynamic and Laplace pressures. Their study concluded that the minimum velocity of the liquid jet required to fragment a micro-sized droplet is about 40 m/s, which is close to our experimental data considering that our length scale is about 15 times larger.

\subsection{Droplet pinch-off}\label{sect:discussion2}

\begin{figure}
	\centering\includegraphics[width=12cm]{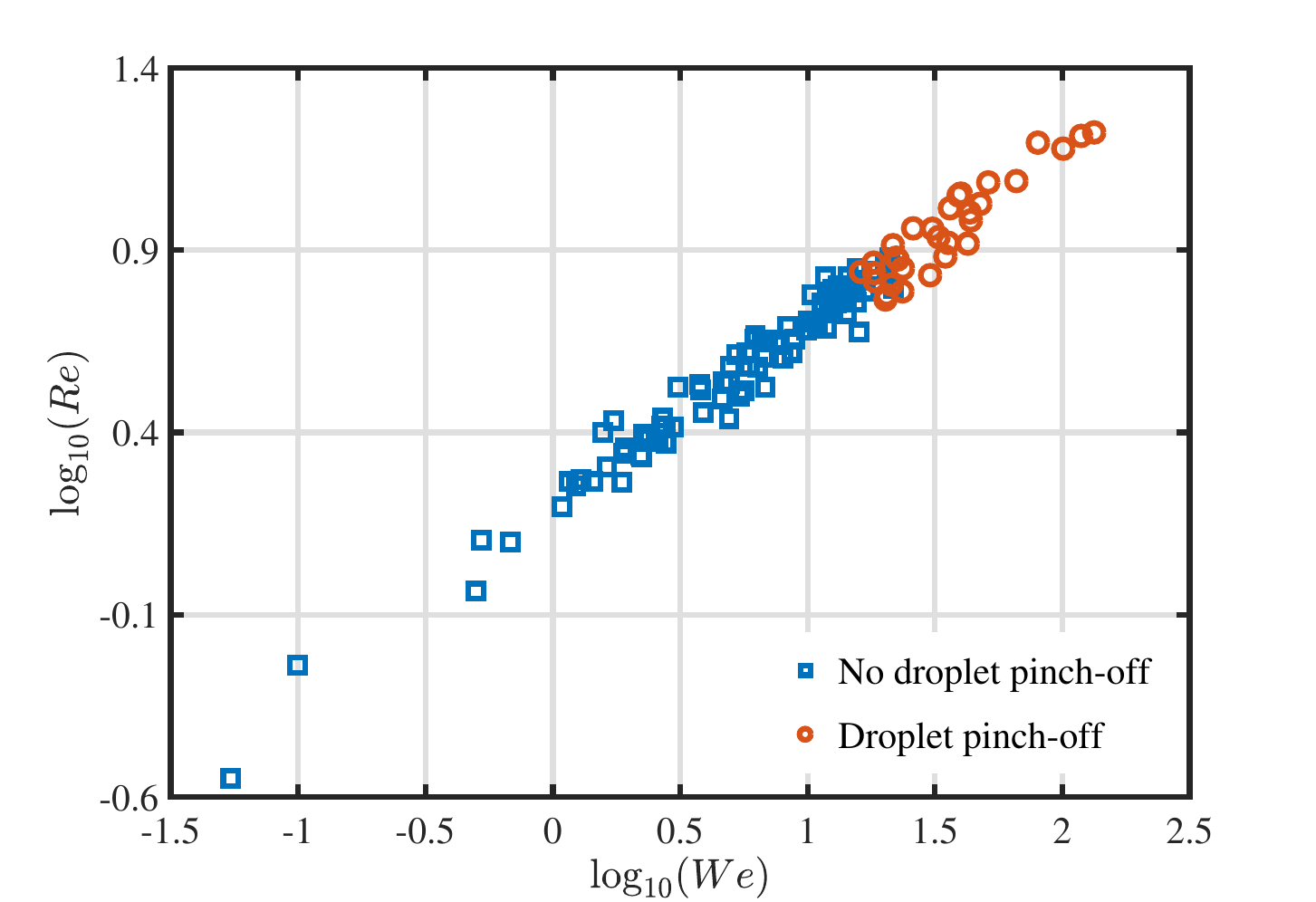}
	\caption{Water-in-oil (W/O) droplet behaviors driven by a migrating vortex ring bubble on a large time scale. The Reynolds and Weber numbers are defined by Equation (\ref{Equation:ReWe}). We identify the critical Weber number denoted as $We_c\approx16$, which demarcates two distinct regimes: (a) the creation of a satellite droplet and (b) the absence of droplet pinch-off.
	}\label{Fig:pinch_off}
\end{figure}

In addition to the jet penetration in the O/W system, another significant mechanism for fluid mixing in the W/O system involves the pinch-off of water droplets. This process occurs on a larger time and length scale compared to jet penetration. Refer to Figure \ref{Fig:water-in-oil} (c) for a representative experiment. As the vortex ring bubble travels over a considerable distance, it induces a global motion in the water droplet. In the context of droplet pinch-off, it is imperative for inertia to overcome surface tension in the presence of viscous dissipation. Consequently, we introduce the local Reynolds and Weber numbers as follows:
\begin{equation}
Re=\frac{\rho_2 R_p \it V_p}{\mu_2},\ \ We=\frac{\rho_1 V_p^2 R_p}{\sigma_2},  \label{Equation:ReWe}
\end{equation}
where $R_p$ represents the radius of the protrusion driven by the bubble and $V_p$ signifies the maximum velocity of the protrusion tip. $R_p$ is approximately equivalent to the radius of the satellite droplet following pinch-off. More than 100 experiments were conducted to establish the droplet breakup criteria. In the phase diagram depicted in Figure \ref{Fig:pinch_off}, we identify the critical Weber number denoted as $We_c$, which demarcates two distinct regimes: (a) the creation of a satellite droplet and (b) the absence of droplet pinch-off. Notably, the critical value $We_c$ is approximately 16, with a corresponding Reynolds number of around 7. This suggests that both surface tension and viscosity play significant roles in this process. Unfortunately, this phenomenon falls beyond the applicability of our BI model. To comprehend this process, we provide an explanation based on the balance between interfacial and kinetic energies. Firstly, the increase in interfacial energy during the droplet pinch-off can be expressed as:
\begin{equation}
\Delta E_i= 4\pi R_p^2 n \sigma_2,  \label{Equation:Ei}
\end{equation}
where $4\pi R_p^2 n$ represents the maximum increment in the interface area during the process, and $n$ is a constant, assuming a value of 1 when the satellite droplet is much smaller than the main droplet. The kinetic energy associated with the migrating remnant bubble can be roughly estimated as:
\begin{equation}
E_k=\frac{1}{2} \rho_1 (\pi R_p^2 l) V_p^2,  \label{Equation:Ekp}
\end{equation}
where $l$ denotes the length of the protrusion jet at the moment when the droplet surface reaches its maximum velocity. Since the vortex ring (bubble) typically carries a portion of liquid that assumes an ellipsoidal shape, with the ratio of semi-minor to semi-major axes typically falling between 1 and 2  \citep{Didden,Dabiri,Sullivan}, we have the relationship $R_p < l < 2R_p$. 

Given that the Reynolds number is less than 10, viscosity leads to considerable energy dissipation. Therefore, the kinetic energy $E_k$ must exceed $\Delta E_i$, resulting in the inequality:
\begin{equation}
We=\frac{\rho_1 V_p^2 R_p}{\sigma_2}>\frac{8nR_p}{l}. \label{Equation:inequ}
\end{equation}

We determined the value of $n$ through our experiments precisely at the point of droplet pinch-off, yielding a range between 1.3 and 1.8. Consequently, the right-hand side of Equation (\ref{Equation:inequ}) spans from 5.2 to 14.4, aligning remarkably well with our experimental observations ($We_c\approx16$).

\section{Summary and conclusions}\label{sect:conclusion}

This paper presents a comprehensive investigation into the dynamics of laser-induced cavitation bubbles within a millimeter-sized droplet suspended in another host fluid, encompassing theoretical, experimental, and numerical aspects. First, we derived a differential equation for spherical bubbles in our particular context and introduced a modified Rayleigh collapse time and natural frequency for the bubble. We systematically assessed the modified Rayleigh factor $\eta$ across a broad parameter space. When the bubble and droplet sizes are comparable, variations in the bubble-to-droplet size ratio $\xi$ significantly impact $\eta$. The analytical estimation of $\eta$ serves as a lower bound for experimental data concerning laser-induced cavitation bubbles. Furthermore, the results from the extended Rayleigh-Plesset equation (Equation \ref{Equation:ERP}) demonstrate excellent alignment with experimental data when the initial parameters are appropriately configured.

We have carried out hundreds of laser-induced cavitation bubble experiments and boundary integral simulations to reveal the dependence of the bubble jetting behaviors and the associated droplet evolution on the governing parameters. A classic scaling \citep{Blake2015,Supponen2016,Han2022} of the nondimensional Kelvin impulse of a bubble $|I_z|$ near a flat fluid-fluid interface with respect to the Atwood number still holds well for the current curved fluid-fluid interface. As $\xi$ increases from a very small value, $|I_z|$ initially increases and then decreases.  Interestingly, $|I_z|$ is maximized at $\xi \approx 0.3$, independent of the density ratio $\alpha$. Remarkably, our BI model accurately capture the evolution of a nonspherical jetting bubble and the associated droplet dynamics. 

The bubbles dynamics in the $\alpha-\gamma$ parameter space was also discussed with a typical values of $\xi=0.6$. Increasing $\alpha$ leads to an increase in jet volume and a decrease in jet impact velocity $V_{jet}$. Increasing $\gamma$ within the range of $\left[ 0.2,\ 1\right] $ results in a reduction in the jet volume and an increase in $V_{jet}$. The kinetic energy of the liquid jet $E_{jet}$ doesn't vary monotonically with $\gamma$. $E_{jet}$ reaches its maximum value at an optimal standoff parameter $\gamma_o$, which increases with the density ratio $\alpha$.

Our experiments have unveiled two distinct mechanisms governing fluid mixing within the O/W and W/O systems, respectively. Firstly, in the O/W system, a remarkably slender liquid jet forms around the end of the first bubble cycle. The maximum velocity of jet impact exceeds 300 m/s. This jet can effectively penetrate the droplet interface when the standoff parameter $\gamma$ is relatively small, specifically $\gamma\lesssim0.88$. A more direct quantification of the jet penetration process is attainable by observing the maximum velocity at the point of impact on the droplet surface, denoted as $V_{d}$. We have determined that the minimum threshold for $V_{d}$ required for successful jet penetration is approximately 31 m/s at the sub-millimeter scale within the context of this paper. Secondly, in the scenario involving W/O droplets, the bubble traverses the interior of the droplet, inducing a global motion within the water droplet. When the inertia of the traveling vortex ring bubble overcomes the surface tension of the droplet, even in the presence of viscous dissipation, the droplet undergoes a division into two daughter droplets. We have established that the critical value, denoted as $We_c$, is approximately 16. This phenomenon is elucidated by considering the equilibrium between interfacial and kinetic energies.

~\\
\noindent\textbf{Acknowledgements.} The authors thank Han Wu and Ying Chen from HEU for the preparation of the experiment.
~\\
\noindent\textbf{Funding.} This work is supported by the National Natural Science Foundation of China (Nos. 12072087, 12372239, 52088102), the Heilongjiang Provincial Natural Science Foundation of China (YQ2022E017), and the Xplore Prize. 

~\\
\noindent\textbf{Declaration of interests.} The authors report no conflict of interest.

\appendix

\begin{figure}
	\centering\includegraphics[width=12cm]{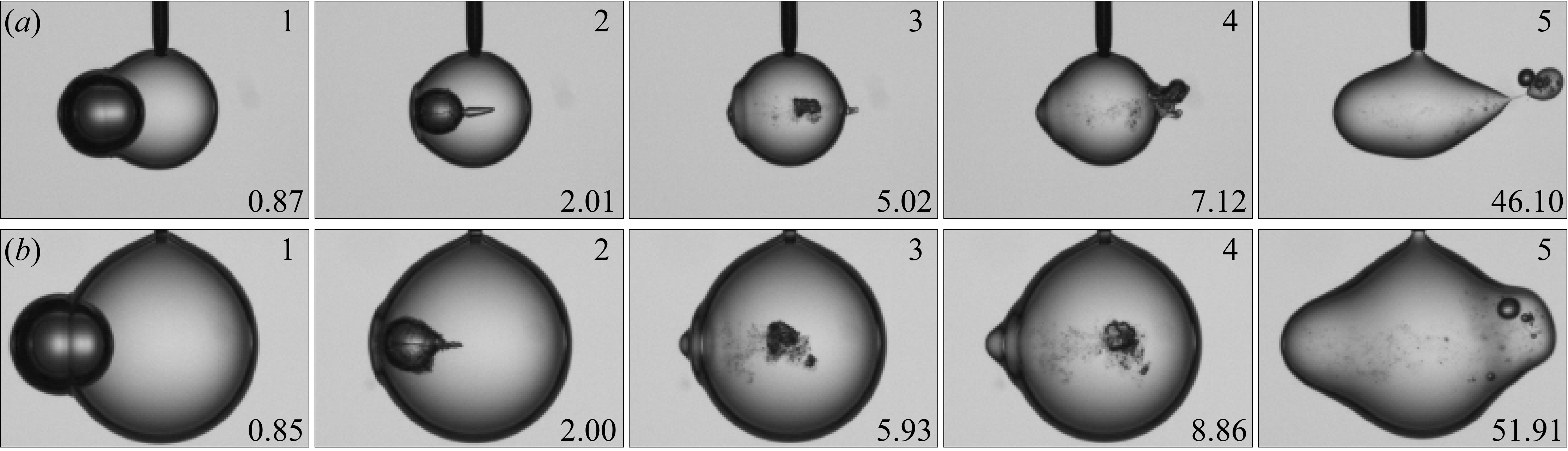}
	\caption{Experimental observation of a laser-induced cavitation bubble at the interface of a water droplet suspended in sunflower oil. $(a)$ Parameters: $R_{d, min} = 0.87$ mm, $R_{b, max } = 0.69$ mm, $\gamma=0$,  and the time scale for nondimensionalization is 69.87 \textmu s. $(b)$ Parameters: $R_{d, min} = 1.50$ mm, $R_{b, max } = 0.78$ mm, $\gamma=0$, and the time scale for nondimensionalization is 78.99 \textmu s. The nondimensional times are indicated at the lower right corner of each frame. The horizontal width of each frame is 4.6 mm.}\label{Fig:protrusion}
\end{figure}

\section{Bubble initiation at the surface of a W/O droplet} \label{sect:appen2}

Figure \ref{Fig:protrusion} presents experimental observations of a particular scenario where a bubble is initiated at the surface of W/O droplets ($\gamma = 0$). Panels $(a)$ and $(b)$ illustrate the influence of interface curvature on bubble dynamics and the evolution of the droplet surface. It is evident that both bubbles in the two experiments produce a high-speed liquid jet that directed from left to right. In the case of the smaller droplet, the bubble has the potential to cause the pinch-off of the droplet. Conversely, in the case of the larger droplet, the bubble leads to substantial deformation on the right side of the interface; however, it does not result in the pinch-off of the droplet. Regarding the left side of the droplet, it is worth noting that the protrusion is less pronounced in the case of the smaller droplet. The phenomenon of interface jet pinch-off, which was observed during the interaction between an electric discharge cavitation bubble and an initially flat water-oil interface \citep{Han2022}, is not observed in the present system.

\section{Sensitivity study on the initial bubble pressure} \label{sect:appen}

When comparing experimental data with the theoretical/numerical model, it is crucial to configure the initial parameters appropriately. The relationship between the nondimensional initial bubble radius $R_0$, initial internal pressure $\varepsilon=P_0/P_\infty$ and the ratio of the specific heats $\lambda$ \citep{Klaseboer04} can be derived from the energy conservation law :

\begin{equation} 
\frac{\varepsilon}{\lambda-1}(R_0^{3\lambda}-R_0^3)=R_0^3-1.\label{Equation:R0}
\end{equation}

\begin{figure}
	\centering\includegraphics[width=12cm]{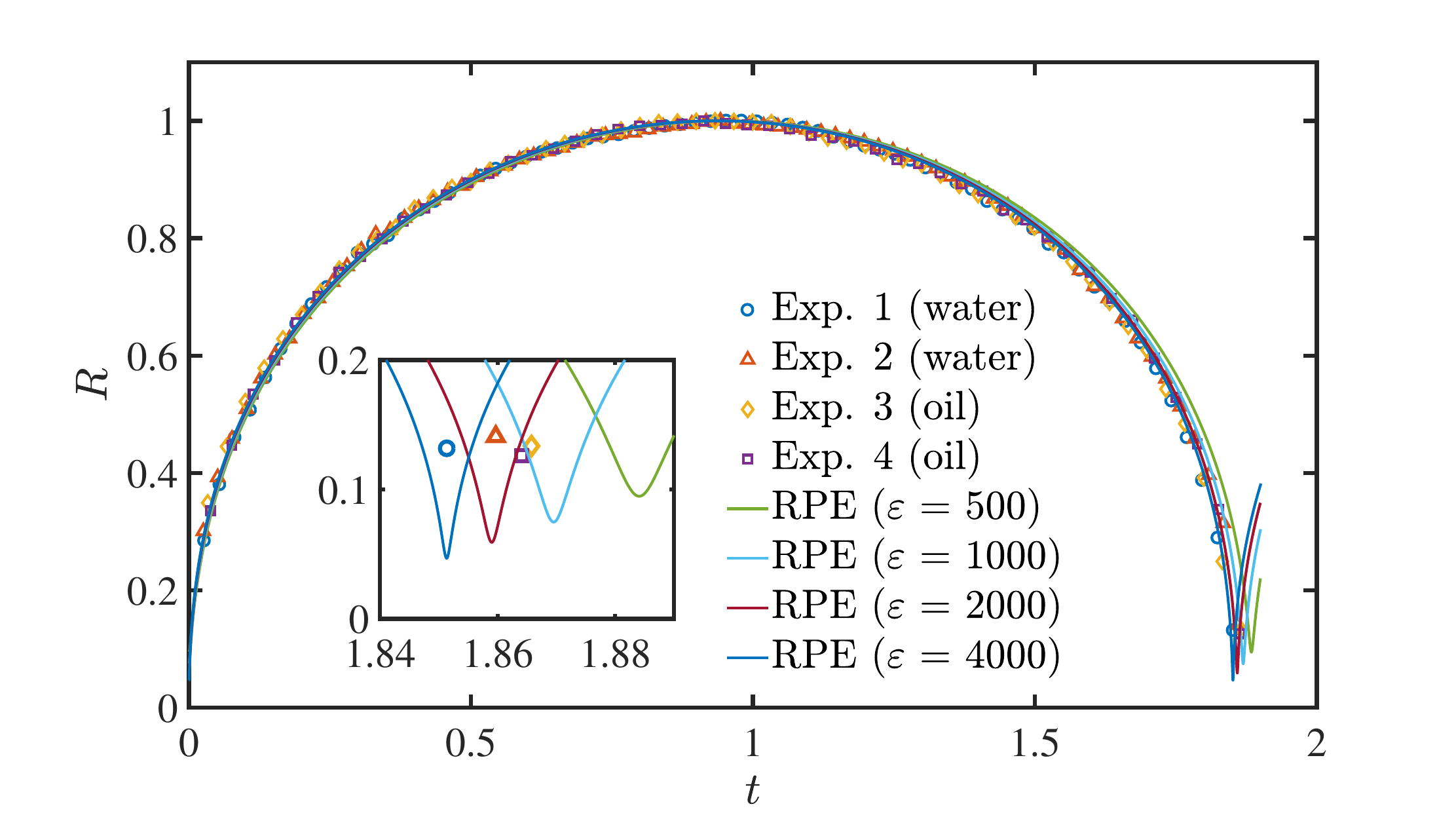}
	\caption{Comparison of experimental data with theoretical results for the nondimensional bubble radius evolution in an unbounded environment. The first two experiments were conducted in water, while the remaining two were carried out in sunflower oil. We present four theoretical results corresponding to different $\varepsilon$ values. The inset shows the moment around the minimum volume of the bubble. The time and length are scaled by $R_{b, max}\it \sqrt{\rho/P_\infty}$ and $R_{b, max}$, respectively.}\label{Fig:comparison-rp-exp}
\end{figure}

Generally, the polytropic exponent $\lambda$ is set as 1.4. We need to adjust only the strength parameter $\varepsilon$ to match the experimental data, while the initial bubble radius is calculated Equation (\ref{Equation:R0}). As shown in Figure \ref{Fig:comparison-rp-exp}, the data obtained from four experiments are presented and compared against the theoretical results obtained from the Rayleigh-Plesset equation (RPE), including bubble initiation in water and sunflower oil. Four different $\varepsilon$ are chosen, namely, 500, 1000, 2000, and 4000, respectively. Since we normalized the time and length with  $R_{b, max}\it \sqrt{\rho/P_\infty}$ and $R_{b, max}$, the experimental data collapse together quite well. The RPE reproduces the experiments quite well except for a slight deviation of the bubble oscillation period $T_{osc}$. The overall bubble dynamics is insensitive to the choice of $\varepsilon$. As inferred from the inset, a satisfactory result can be obtained if $\varepsilon$ is set as 2000.

\bibliographystyle{jfm}

\bibliography{sample}

\begin{thebibliography}{60}
\expandafter\ifx\csname natexlab\endcsname\relax\def\natexlab#1{#1}\fi
\def\au#1{#1} \def\ed#1{#1} \def\yr#1{#1}\def\at#1{#1}\def\jt#1{\textit{#1}}
  \def\bt#1{#1}\def\bvol#1{\textbf{#1}} \def\vol#1{#1} \def\pg#1{#1}
  \def\publ#1{#1}\def\arxiv#1{#1}\def\org#1{#1}\def\st#1{\textit{#1}}

\bibitem[Banine {\em et~al.\/}(2011)Banine, Koshelev \& Swinkels]{Banine}
{\sc \au{Banine, V.~Y.}, \au{Koshelev, K.~N.} \& \au{Swinkels, G.}} \yr{2011}
  \at{Physical processes in {EUV} sources for microlithography}.  \jt{Journal
  of Physics D: Applied Physics}  \bvol{44}~(25),  \pg{253001}.

\bibitem[Bempedelis {\em et~al.\/}(2021)Bempedelis, Zhou, Andersson \&
  Ventikos]{Bempedelis2021}
{\sc \au{Bempedelis, N.}, \au{Zhou, J.}, \au{Andersson, M.} \& \au{Ventikos,
  Y.}} \yr{2021}  \at{Numerical and experimental investigation into the
  dynamics of a bubble-free-surface system}.  \jt{Physical Review Fluids}
  \bvol{6}~(1),  \pg{013606}.

\bibitem[Blake {\em et~al.\/}(2015)Blake, Leppinen \& Wang]{Blake2015}
{\sc \au{Blake, J.~R.}, \au{Leppinen, D.~M.} \& \au{Wang, Q.}} \yr{2015}
  \at{Cavitation and bubble dynamics: the {K}elvin impulse and its
  applications}.  \jt{Interface Focus}  \bvol{5}~(5),  \pg{20150017}.

\bibitem[Brujan {\em et~al.\/}(2001)Brujan, Nahen, Schmidt \& Vogel]{Brujan01}
{\sc \au{Brujan, E.~A.}, \au{Nahen, K.}, \au{Schmidt, P.} \& \au{Vogel, A.}}
  \yr{2001}  \at{Dynamics of laser-induced cavitation bubbles near an elastic
  boundary}.  \jt{Journal of Fluid Mechanics}  \bvol{433},  \pg{251--281}.

\bibitem[Brujan {\em et~al.\/}(2018)Brujan, Noda, Ishigami, Ogasawara \&
  Takahira]{Brujan2018}
{\sc \au{Brujan, E.-A.}, \au{Noda, T.}, \au{Ishigami, A.}, \au{Ogasawara, T.}
  \& \au{Takahira, H.}} \yr{2018}  \at{Dynamics of laser-induced cavitation
  bubbles near two perpendicular rigid walls}.  \jt{Journal of Fluid Mechanics}
   \bvol{841},  \pg{28--49}.

\bibitem[Cerbus {\em et~al.\/}(2022)Cerbus, Chraibi, Tondusson, Petit, Soto,
  Devillard, Delville \& Kellay]{Cerbus2022}
{\sc \au{Cerbus, R.~T.}, \au{Chraibi, H.}, \au{Tondusson, M.}, \au{Petit, S.},
  \au{Soto, D.}, \au{Devillard, R.}, \au{Delville, J.~P.} \& \au{Kellay, H.}}
  \yr{2022}  \at{Experimental and numerical study of laser-induced secondary
  jetting}.  \jt{Journal of Fluid Mechanics}  \bvol{934},  \pg{A14}.

\bibitem[Chahine \& Bovis(1980)]{Chahine1980}
{\sc \au{Chahine, G.~L.} \& \au{Bovis, A.}} \yr{1980} Oscillation and collapse
  of a cavitation bubble in the vicinity of a two-liquid interface.  \bt{In
  {\em Cavitation and Inhomogeneities in Underwater Acoustics\/} (ed.
  \ed{Werner Lauterborn})},  \pg{pp. 23--29}.  \publ{Springer Berlin
  Heidelberg}.

\bibitem[Church(1995)]{Church}
{\sc \au{Church, C.~C.}} \yr{1995}  \at{The effects of an elastic solid surface
  layer on the radial pulsations of gas bubbles}.  \jt{The Journal of the
  Acoustical Society of America}  \bvol{97}~(3),  \pg{1510--1521}.

\bibitem[Curtiss {\em et~al.\/}(2013)Curtiss, Leppinen, Wang \&
  Blake]{Curtiss2013}
{\sc \au{Curtiss, G.~A.}, \au{Leppinen, D.~M.}, \au{Wang, Q.~X.} \& \au{Blake,
  J.~R.}} \yr{2013}  \at{Ultrasonic cavitation near a tissue layer}.
  \jt{Journal of Fluid Mechanics}  \bvol{730},  \pg{245--272}.

\bibitem[Dabiri \& Gharib(2004)]{Dabiri}
{\sc \au{Dabiri, J.~O.} \& \au{Gharib, M.}} \yr{2004}  \at{Fluid entrainment by
  isolated vortex rings}.  \jt{Journal of Fluid Mechanics}  \bvol{511},
  \pg{311--331}.

\bibitem[Didden(1979)]{Didden}
{\sc \au{Didden, N.}} \yr{1979}  \at{On the formation of vortex rings:
  Rolling-up and production of circulation}.  \jt{Journal of Applied
  Mathematics and Physics}  \bvol{30},  \pg{101--116}.

\bibitem[Gonzalez~Avila \& Ohl(2016)]{Gonzalez}
{\sc \au{Gonzalez~Avila, S.~R.} \& \au{Ohl, C.-D.}} \yr{2016}
  \at{Fragmentation of acoustically levitating droplets by laser-induced
  cavitation bubbles}.  \jt{Journal of Fluid Mechanics}  \bvol{805},
  \pg{551--576}.

\bibitem[Han {\em et~al.\/}(2023{\natexlab{{\em a\/}}})Han, Liu, Ren, Sun,
  Tagawa, Zeng \& Zuo]{HanH2023}
{\sc \au{Han, H.}, \au{Liu, S.}, \au{Ren, Z.}, \au{Sun, C.}, \au{Tagawa, Y.},
  \au{Zeng, H.} \& \au{Zuo, Z.}} \yr{2023{\natexlab{{\em a\/}}}}
  \at{Interactions of a collapsing laser-induced cavitation bubble with a
  hemispherical droplet attached to a rigid boundary}.  \jt{Journal of Fluid
  Mechanics}  \bvol{976},  \pg{A11}.

\bibitem[Han {\em et~al.\/}(2023{\natexlab{{\em b\/}}})Han, Zhang, Yang, Han \&
  Li]{HanL}
{\sc \au{Han, L.}, \au{Zhang, T.}, \au{Yang, D.}, \au{Han, R.} \& \au{Li, S.}}
  \yr{2023{\natexlab{{\em b\/}}}}  \at{Comparison of vortex cut and vortex ring
  models for toroidal bubble dynamics in underwater explosions}.  \jt{Fluids}
  \bvol{8},  \pg{131}.

\bibitem[Han {\em et~al.\/}(2022)Han, Zhang, Tan \& Li]{Han2022}
{\sc \au{Han, R.}, \au{Zhang, A.~M.}, \au{Tan, S.} \& \au{Li, S.}} \yr{2022}
  \at{Interaction of cavitation bubbles with the interface of two immiscible
  fluids on multiple time scales}.  \jt{Journal of Fluid Mechanics}
  \bvol{932},  \pg{A8}.

\bibitem[Hsiao {\em et~al.\/}(2014)Hsiao, Jayaprakash, Kapahi, Choi \&
  Chahine]{Hsiao2014}
{\sc \au{Hsiao, C.~T.}, \au{Jayaprakash, A.}, \au{Kapahi, A.}, \au{Choi, J.~K.}
  \& \au{Chahine, G.~L.}} \yr{2014}  \at{Modelling of material pitting from
  cavitation bubble collapse}.  \jt{Journal of Fluid Mechanics}  \bvol{755},
  \pg{142--175}.

\bibitem[Janzen {\em et~al.\/}(2005)Janzen, Fleige, Noll, Schwenke, Lahmann,
  Knoth, Beaven, Jantzen, Oest \& Koke]{Janzen}
{\sc \au{Janzen, C.}, \au{Fleige, R.}, \au{Noll, R.}, \au{Schwenke, H.},
  \au{Lahmann, W.}, \au{Knoth, J.}, \au{Beaven, P.}, \au{Jantzen, E.},
  \au{Oest, A.} \& \au{Koke, P.}} \yr{2005}  \at{Analysis of small droplets
  with a new detector for liquid chromatography based on laser-induced
  breakdown spectroscopy}.  \jt{Spectrochimica Acta Part B: Atomic
  Spectroscopy}  \bvol{60}~(7),  \pg{993--1001}.

\bibitem[Kang \& Cho(2019)]{Kang2019}
{\sc \au{Kang, Y.~J.} \& \au{Cho, Y.}} \yr{2019}  \at{Gravity–capillary
  jet-like surface waves generated by an underwater bubble}.  \jt{Journal of
  Fluid Mechanics}  \bvol{866},  \pg{841--864}.

\bibitem[Kannan {\em et~al.\/}(2020)Kannan, Balusamy, Karri \&
  Sahu]{Kannan2020}
{\sc \au{Kannan, Y.~S.}, \au{Balusamy, S.}, \au{Karri, B.} \& \au{Sahu, K.~C.}}
  \yr{2020}  \at{Effect of viscosity on the volumetric oscillations of a
  non-equilibrium bubble in free-field and near a free-surface}.
  \jt{Experimental Thermal and Fluid Science}  \bvol{116},  \pg{110113}.

\bibitem[Klaseboer \& Khoo(2004{\natexlab{{\em a\/}}})]{Klaseboer-CM2004}
{\sc \au{Klaseboer, E.} \& \au{Khoo, B.~C.}} \yr{2004{\natexlab{{\em a\/}}}}
  \at{Boundary integral equations as applied to an oscillating bubble near a
  fluid-fluid interface}.  \jt{Computational Mechanics}  \bvol{33}~(2),
  \pg{129--138}.

\bibitem[Klaseboer \& Khoo(2004{\natexlab{{\em b\/}}})]{Klaseboer04}
{\sc \au{Klaseboer, E} \& \au{Khoo, B.~C.}} \yr{2004{\natexlab{{\em b\/}}}}
  \at{An oscillating bubble near an elastic material}.  \jt{Journal of Applied
  Physics}  \bvol{96}~(10),  \pg{5808--5818}.

\bibitem[Klein {\em et~al.\/}(2020)Klein, Kurilovich, Lhuissier, Versolato,
  Lohse, Villermaux \& Gelderblom]{Klein2020}
{\sc \au{Klein, A.~L.}, \au{Kurilovich, D.}, \au{Lhuissier, H.}, \au{Versolato,
  O.~O.}, \au{Lohse, D.}, \au{Villermaux, E.} \& \au{Gelderblom, H.}} \yr{2020}
   \at{Drop fragmentation by laser-pulse impact}.  \jt{Journal of Fluid
  Mechanics}  \bvol{893},  \pg{A7}.

\bibitem[Li {\em et~al.\/}(2020{\natexlab{{\em a\/}}})Li, van~der Meer, Zhang,
  Prosperetti \& Lohse]{Li2020}
{\sc \au{Li, S.}, \au{van~der Meer, D.}, \au{Zhang, A.-M.}, \au{Prosperetti,
  A.} \& \au{Lohse, D.}} \yr{2020{\natexlab{{\em a\/}}}}  \at{Modelling large
  scale airgun-bubble dynamics with highly non-spherical features}.
  \jt{International Journal of Multiphase Flow}  \bvol{122},  \pg{103143}.

\bibitem[Li {\em et~al.\/}(2020{\natexlab{{\em b\/}}})Li, Prosperetti \&
  van~der Meer]{LiS2020}
{\sc \au{Li, S.}, \au{Prosperetti, A.} \& \au{van~der Meer, D.}}
  \yr{2020{\natexlab{{\em b\/}}}}  \at{Dynamics of a toroidal bubble on a
  cylinder surface with an application to geophysical exploration}.
  \jt{International Journal of Multiphase Flow}  \bvol{129},  \pg{103335}.

\bibitem[Li {\em et~al.\/}(2023)Li, Zhang \& Han]{Li_jcp}
{\sc \au{Li, S.}, \au{Zhang, A.-M.} \& \au{Han, R.}} \yr{2023}  \at{3{D} model
  for inertial cavitation bubble dynamics in binary immiscible fluids}.
  \jt{Journal of Computational Physics}  \bvol{494},  \pg{112508}.

\bibitem[Liang {\em et~al.\/}(2020)Liang, Jiang, Wen \& Liu]{Liang2020}
{\sc \au{Liang, Y.}, \au{Jiang, Y.}, \au{Wen, C.-Y.} \& \au{Liu, Y.}} \yr{2020}
   \at{Interaction of a planar shock wave and a water droplet embedded with a
  vapour cavity}.  \jt{Journal of Fluid Mechanics}  \bvol{885},  \pg{R6}.

\bibitem[{Lord Rayleigh}(1917)]{Rayleigh}
{\sc \au{{Lord Rayleigh}}} \yr{1917}  \at{{VIII}. {O}n the pressure developed
  in a liquid during the collapse of a spherical cavity}.  \jt{The London,
  Edinburgh, and Dublin Philosophical Magazine and Journal of Science}
  \bvol{34}~(200),  \pg{94--98}.

\bibitem[Luo {\em et~al.\/}(2021)Luo, Xu \& Khoo]{Luo2021}
{\sc \au{Luo, J.}, \au{Xu, W.} \& \au{Khoo, B.~C.}} \yr{2021}
  \at{Stratification effect of air bubble on the shock wave from the collapse
  of cavitation bubble}.  \jt{Journal of Fluid Mechanics}  \bvol{919},
  \pg{A16}.

\bibitem[Minnaert(1933)]{Minnaert}
{\sc \au{Minnaert, M.}} \yr{1933}  \at{{XVI}. on musical air-bubbles and the
  sounds of running water}.  \jt{The London, Edinburgh, and Dublin
  Philosophical Magazine and Journal of Science}  \bvol{16}~(104),
  \pg{235--248}.

\bibitem[Minsier {\em et~al.\/}(2009)Minsier, De~Wilde \& Proost]{Minsier}
{\sc \au{Minsier, V.}, \au{De~Wilde, J.} \& \au{Proost, J.}} \yr{2009}
  \at{Simulation of the effect of viscosity on jet penetration into a single
  cavitating bubble}.  \jt{Journal of Applied Physics}  \bvol{106}~(8),
  \pg{084906}.

\bibitem[Obreschkow {\em et~al.\/}(2006)Obreschkow, Kobel, Dorsaz, De~Bosset,
  Nicollier \& Farhat]{Obreschkow}
{\sc \au{Obreschkow, D.}, \au{Kobel, P.}, \au{Dorsaz, N.}, \au{De~Bosset, A.},
  \au{Nicollier, C.} \& \au{Farhat, M.}} \yr{2006}  \at{Cavitation bubble
  dynamics inside liquid drops in microgravity}.  \jt{Physical Review Letters}
  \bvol{97}~(9),  \pg{094502}.

\bibitem[Orthaber {\em et~al.\/}(2020)Orthaber, Zevnik, Petkovšek \&
  Dular]{Orthaber}
{\sc \au{Orthaber, U.}, \au{Zevnik, J.}, \au{Petkovšek, R.} \& \au{Dular, M.}}
  \yr{2020}  \at{Cavitation bubble collapse in a vicinity of a liquid-liquid
  interface – basic research into emulsification process}.  \jt{Ultrasonics
  Sonochemistry}  \bvol{68},  \pg{105224}.

\bibitem[Pearson {\em et~al.\/}(2004)Pearson, Blake \& Otto]{Pearson}
{\sc \au{Pearson, A.}, \au{Blake, J.R.} \& \au{Otto, S.R.}} \yr{2004}  \at{Jets
  in bubbles}.  \jt{Journal of Engineering Mathematics}  \bvol{48}~(3-4),
  \pg{391--412}.

\bibitem[Perdih {\em et~al.\/}(2019)Perdih, Zupanc \& Dular]{Stepisnik}
{\sc \au{Perdih, S.~T.}, \au{Zupanc, M.} \& \au{Dular, M.}} \yr{2019}
  \at{Revision of the mechanisms behind oil-water (o/w) emulsion preparation by
  ultrasound and cavitation}.  \jt{Ultrasonics Sonochemistry}  \bvol{51},
  \pg{298--304}.

\bibitem[Plesset(1954)]{Plesset1954}
{\sc \au{Plesset, M.~S.}} \yr{1954}  \at{On the stability of fluid flows with
  spherical symmetry}.  \jt{Journal of Applied Physics}  \bvol{25}~(1),
  \pg{96--98}.

\bibitem[Popinet \& Zaleski(2002)]{Popinet}
{\sc \au{Popinet, S.} \& \au{Zaleski, S.}} \yr{2002}  \at{Bubble collapse near
  a solid boundary: a numerical study of the influence of viscosity}.
  \jt{Journal of Fluid Mechanics}  \bvol{464},  \pg{137--163}.

\bibitem[Poulain {\em et~al.\/}(2015)Poulain, Guenoun, Gart, Crowe \&
  Jung]{Poulain15}
{\sc \au{Poulain, S.}, \au{Guenoun, G.}, \au{Gart, S.}, \au{Crowe, W.} \&
  \au{Jung, S.}} \yr{2015}  \at{Particle motion induced by bubble cavitation}.
  \jt{Physical Review Letters}  \bvol{114}~(21),  \pg{214501}.

\bibitem[Prosperetti(1977)]{Prosperetti1977}
{\sc \au{Prosperetti, A.}} \yr{1977}  \at{Viscous effects on perturbed
  spherical flows}.  \jt{Quarterly of Applied Mathematics}  \bvol{34}~(4),
  \pg{339--352}.

\bibitem[Raman {\em et~al.\/}(2022{\natexlab{{\em a\/}}})Raman, Rosselló,
  Reese \& Ohl]{Raman2022JFM}
{\sc \au{Raman, K.~A.}, \au{Rosselló, J.M.}, \au{Reese, H.} \& \au{Ohl,
  C.-D.}} \yr{2022{\natexlab{{\em a\/}}}}  \at{Microemulsification from single
  laser-induced cavitation bubbles}.  \jt{Journal of Fluid Mechanics}
  \bvol{953},  \pg{A27}.

\bibitem[Raman {\em et~al.\/}(2022{\natexlab{{\em b\/}}})Raman, Rosselló \&
  Ohl]{Raman2022}
{\sc \au{Raman, K.~A.}, \au{Rosselló, J.~M.} \& \au{Ohl, C.-D.}}
  \yr{2022{\natexlab{{\em b\/}}}}  \at{Cavitation induced oil-in-water
  emulsification pathways using a single laser-induced bubble}.  \jt{Applied
  Physics Letters}  \bvol{121}~(19),  \pg{194103}.

\bibitem[Ren {\em et~al.\/}(2022)Ren, Zuo, Wu \& Liu]{Ren2022}
{\sc \au{Ren, Z.}, \au{Zuo, Z.}, \au{Wu, S.} \& \au{Liu, S.}} \yr{2022}
  \at{Particulate projectiles driven by cavitation bubbles}.  \jt{Physical
  Review Letters}  \bvol{128}~(4),  \pg{044501}.

\bibitem[Rosselló {\em et~al.\/}(2023)Rosselló, Reese, Raman \&
  Ohl]{Rossello}
{\sc \au{Rosselló, J.~M.}, \au{Reese, H.}, \au{Raman, K.~A.} \& \au{Ohl,
  C.-D.}} \yr{2023}  \at{Bubble nucleation and jetting inside a millimetric
  droplet}.  \jt{Journal of Fluid Mechanics}  \bvol{968},  \pg{A19}.

\bibitem[Saade {\em et~al.\/}(2021)Saade, Jalaal, Prosperetti \&
  Lohse]{saade2021}
{\sc \au{Saade, Y.}, \au{Jalaal, M.}, \au{Prosperetti, A.} \& \au{Lohse, D.}}
  \yr{2021}  \at{Crown formation from a cavitating bubble close to a free
  surface}.  \jt{Journal of Fluid Mechanics}  \bvol{926},  \pg{A5}.

\bibitem[Saini {\em et~al.\/}(2022)Saini, Tanne, Arrigoni, Zaleski \&
  Fuster]{Saini2022}
{\sc \au{Saini, M.}, \au{Tanne, E.}, \au{Arrigoni, M.}, \au{Zaleski, S.} \&
  \au{Fuster, D.}} \yr{2022}  \at{On the dynamics of a collapsing bubble in
  contact with a rigid wall}.  \jt{Journal of Fluid Mechanics}  \bvol{948},
  \pg{A45}.

\bibitem[Sullivan {\em et~al.\/}(2008)Sullivan, Niemela, Hershberger, Bolster
  \& Donnelly]{Sullivan}
{\sc \au{Sullivan, I.~S.}, \au{Niemela, J.~J.}, \au{Hershberger, R.~E.},
  \au{Bolster, D.} \& \au{Donnelly, R.~J.}} \yr{2008}  \at{Dynamics of thin
  vortex rings}.  \jt{Journal of Fluid Mechanics}  \bvol{609},  \pg{319--347}.

\bibitem[Supponen {\em et~al.\/}(2016)Supponen, Obreschkow, Tinguely, Kobel,
  Dorsaz \& Farhat]{Supponen2016}
{\sc \au{Supponen, O.}, \au{Obreschkow, D.}, \au{Tinguely, M.}, \au{Kobel, P.},
  \au{Dorsaz, N.} \& \au{Farhat, M.}} \yr{2016}  \at{Scaling laws for jets of
  single cavitation bubbles}.  \jt{Journal of Fluid Mechanics}  \bvol{802},
  \pg{263--293}.

\bibitem[Tomita \& Sato(2017)]{Tomita17}
{\sc \au{Tomita, Y.} \& \au{Sato, K.}} \yr{2017}  \at{Pulsed jets driven by two
  interacting cavitation bubbles produced at different times}.  \jt{Journal of
  Fluid Mechanics}  \bvol{819},  \pg{465--493}.

\bibitem[Udepurkar {\em et~al.\/}(2023)Udepurkar, Clasen \& Kuhn]{Udepurkar}
{\sc \au{Udepurkar, A.~P.}, \au{Clasen, C.} \& \au{Kuhn, S.}} \yr{2023}
  \at{Emulsification mechanism in an ultrasonic microreactor: Influence of
  surface roughness and ultrasound frequency}.  \jt{Ultrasonics Sonochemistry}
  \bvol{94},  \pg{106323}.

\bibitem[Wang {\em et~al.\/}(2021)Wang, Li, Guo, Wang, Du, Wang, Abe \&
  Huang]{WangJZ2021}
{\sc \au{Wang, J.}, \au{Li, H.}, \au{Guo, W.}, \au{Wang, Z.}, \au{Du, T.},
  \au{Wang, Y.}, \au{Abe, A.} \& \au{Huang, C.}} \yr{2021}
  \at{Rayleigh–{T}aylor instability of cylindrical water droplet induced by
  laser-produced cavitation bubble}.  \jt{Journal of Fluid Mechanics}
  \bvol{919},  \pg{A42}.

\bibitem[Wang {\em et~al.\/}(2022)Wang, Liu, Corbett \& Smith]{WangQ2022}
{\sc \au{Wang, Q.}, \au{Liu, W.}, \au{Corbett, C.} \& \au{Smith, W.~R.}}
  \yr{2022}  \at{Microbubble dynamics in a viscous compressible liquid subject
  to ultrasound}.  \jt{Physics of Fluids}  \bvol{34}~(1),  \pg{012105}.

\bibitem[Wang {\em et~al.\/}(1996)Wang, Yeo, Khoo \& Lam]{WangQX1996}
{\sc \au{Wang, Q.~X.}, \au{Yeo, K.~S.}, \au{Khoo, B.~C.} \& \au{Lam, K.~Y.}}
  \yr{1996}  \at{Nonlinear interaction between gas bubble and free surface}.
  \jt{Computers \& Fluids}  \bvol{25}~(7),  \pg{607--628}.

\bibitem[Wu {\em et~al.\/}(2021)Wu, Eskin, Priyadarshi, Subroto, Tzanakis \&
  Zhai]{Wu2023}
{\sc \au{Wu, W.~H.}, \au{Eskin, D.~G.}, \au{Priyadarshi, A.}, \au{Subroto, T.},
  \au{Tzanakis, I.} \& \au{Zhai, W.}} \yr{2021}  \at{New insights into the
  mechanisms of ultrasonic emulsification in the oil–water system and the
  role of gas bubbles}.  \jt{Ultrasonics Sonochemistry}  \bvol{73},
  \pg{105501}.

\bibitem[Yamamoto {\em et~al.\/}(2021)Yamamoto, Matsutaka \&
  Komarov]{Yamamoto2021}
{\sc \au{Yamamoto, T.}, \au{Matsutaka, R.} \& \au{Komarov, S.~V.}} \yr{2021}
  \at{High-speed imaging of ultrasonic emulsification using a water-gallium
  system}.  \jt{Ultrasonics Sonochemistry}  \bvol{71},  \pg{105387}.

\bibitem[Yi {\em et~al.\/}(2021)Yi, Li, Jiang, Lohse, Sun \& Mathai]{Yi2021}
{\sc \au{Yi, L.}, \au{Li, S.}, \au{Jiang, H.}, \au{Lohse, D.}, \au{Sun, C.} \&
  \au{Mathai, V.}} \yr{2021}  \at{Water entry of spheres into a rotating
  liquid}.  \jt{Journal of Fluid Mechanics}  \bvol{912},  \pg{R1}.

\bibitem[Zeng {\em et~al.\/}(2022)Zeng, An \& Ohl]{ZengQ2022}
{\sc \au{Zeng, Q.}, \au{An, H.} \& \au{Ohl, C.-D.}} \yr{2022}  \at{Wall shear
  stress from jetting cavitation bubbles: influence of the stand-off distance
  and liquid viscosity}.  \jt{Journal of Fluid Mechanics}  \bvol{932},
  \pg{A14}.

\bibitem[Zeng {\em et~al.\/}(2020)Zeng, Gonzalez-Avila \& Ohl]{Zeng2020}
{\sc \au{Zeng, Q.}, \au{Gonzalez-Avila, S.~R.} \& \au{Ohl, C.-D.}} \yr{2020}
  \at{Splitting and jetting of cavitation bubbles in thin gaps}.  \jt{Journal
  of Fluid Mechanics}  \bvol{896},  \pg{A28}.

\bibitem[Zeng {\em et~al.\/}(2018)Zeng, Gonzalez-Avila, Voorde \&
  Ohl]{zeng-drop}
{\sc \au{Zeng, Q.}, \au{Gonzalez-Avila, S.~R.}, \au{Voorde, S.~T.} \& \au{Ohl,
  C.-D.}} \yr{2018}  \at{Jetting of viscous droplets from cavitation-induced
  {R}ayleigh–{T}aylor instability}.  \jt{Journal of Fluid Mechanics}
  \bvol{846},  \pg{916--943}.

\bibitem[Zhang {\em et~al.\/}(2015)Zhang, Li \& Cui]{Zhang2015pof}
{\sc \au{Zhang, A.~M.}, \au{Li, S.} \& \au{Cui, J.}} \yr{2015}  \at{Study on
  splitting of a toroidal bubble near a rigid boundary}.  \jt{Physics of
  Fluids}  \bvol{27}~(6),  \pg{062102}.

\bibitem[Zhang {\em et~al.\/}(2023{\natexlab{{\em a\/}}})Zhang, Li, Cui, Li \&
  Liu]{Zhang2023a}
{\sc \au{Zhang, A.-M.}, \au{Li, S.-M.}, \au{Cui, P.}, \au{Li, S.} \& \au{Liu,
  Y.-L.}} \yr{2023{\natexlab{{\em a\/}}}}  \at{Theoretical study on bubble
  dynamics under hybrid-boundary and multi-bubble conditions using the unified
  equation}.  \jt{Science China Physics, Mechanics \& Astronomy}
  \bvol{66}~(12),  \pg{124711}.

\bibitem[Zhang {\em et~al.\/}(2023{\natexlab{{\em b\/}}})Zhang, Li, Cui, Li \&
  Liu]{Zhang2023}
{\sc \au{Zhang, A.-M.}, \au{Li, S.-M.}, \au{Cui, P.}, \au{Li, S.} \& \au{Liu,
  Y.-L.}} \yr{2023{\natexlab{{\em b\/}}}}  \at{A unified theory for bubble
  dynamics}.  \jt{Physics of Fluids}  \bvol{35}~(3),  \pg{033323}.

\end{thebibliography}



\end{document}